\documentclass[sigconf, nonacm]{acmart}
\usepackage{algorithm}
\usepackage{algorithmicx}
\usepackage{algpseudocodex}
\usepackage{multirow}
\usepackage{subcaption}
\usepackage{xspace}
\usepackage{tablefootnote}
\usepackage{balance}

\newcommand{\oursystem}{SEFRQO\xspace}
\newcommand{\llm}{LLM}
\newcommand{\qgrpo}{Q$_{GRPO}$}
\newcommand{\pghint}{\textit{pg\_hint\_plan}}

\AtBeginDocument{%
  }

\setcopyright{acmlicensed}
\copyrightyear{2018}
\acmYear{2018}
\acmDOI{XXXXXXX.XXXXXXX}
\acmConference[Conference acronym 'XX]{Make sure to enter the correct
  conference title from your rights confirmation email}{June 03--05,
  2018}{Woodstock, NY}
\acmISBN{978-1-4503-XXXX-X/2018/06}

\begin{document}

\title{SEFRQO: A Self-Evolving Fine-Tuned RAG-Based Query Optimizer}

\settopmatter{authorsperrow=4}
\author{Hanwen Liu}
\orcid{0000-0002-5265-9312}
\authornote{Both authors contributed equally to this work.}
\affiliation{
  \institution{University of Southern California}
  \city{Los Angeles}
  \country{USA}}
\email{hanwen\_liu@usc.edu}

\author{Qihan Zhang}
\authornotemark[1]
\orcid{0009-0005-5785-8766}
\affiliation{%
  \institution{University of Southern California}
  \city{Los Angeles}
  \country{USA}}
\email{qihanzha@usc.edu}

\author{Ryan Marcus}
\orcid{0000-0002-1279-1124}
\affiliation{%
  \institution{University of Pennsylvania}
  \city{Philadelphia}
  \country{USA}}
\email{rcmarcus@seas.upenn.edu}

\author{Ibrahim Sabek}
\orcid{0009-0006-2102-5241}
\affiliation{%
  \institution{University of Southern California}
  \city{Los Angeles}
  \country{USA}}
\email{sabek@usc.edu}

\renewcommand{\shortauthors}{Liu et al.}

\begin{abstract}
Query optimization is a crucial problem in database systems that has been studied for decades. Learned query optimizers (LQOs) can improve performance over time by incorporating feedback; however, they suffer from cold‑start issues and often require retraining when workloads shift or schemas change. Recent LLM‑based query optimizers leverage pre‑trained and fine‑tuned LLMs to mitigate these challenges. Nevertheless, they neglect LLMs' in‑context learning and execution records as feedback for continuous evolution.

In this paper, we present {\oursystem}, a \textbf{S}elf-\textbf{E}volving \textbf{F}ine-tuned \textbf{R}AG-based \textbf{Q}uery \textbf{O}ptimizer. {\oursystem} mitigates the cold-start problem of LQOs by continuously learning from execution feedback via a Retrieval-Augmented Generation (RAG) framework. We employ both supervised fine‑tuning and reinforcement fine‑tuning to prepare the LLM to produce syntactically correct and performance efficient query hints. Moreover, {\oursystem} leverages the LLM’s in‑context learning capabilities by dynamically constructing prompts with references to similar queries and the historical execution record of the same query. This self‑evolving paradigm iteratively optimizes the prompt to minimize query execution latency. Evaluations show that {\oursystem} outperforms state‑of‑the‑art LQOs, achieving up to 65.05\% and 93.57\% reductions in query latency on the CEB and Stack workloads, respectively, compared to PostgreSQL. 
\end{abstract}
\maketitle
\section{Introduction}

Query optimization is a fundamental and performance-critical problem in database systems that focuses on translating declarative user queries into efficient query plans. Human experts take decades to design and improve classical query optimizers (e.g., PostgreSQL~\cite{postgres}). To try and accelerate optimizer development and improve query performance, several studies have applied different techniques to query optimization, such as machine learning~\cite{marcus2021bao, marcus2019neo, rtos,yang2022balsa,zhu2023lero,yu2022cost,doshi2023kepler} and large language models~\cite{llm4qo-1,llm4qo-2}.

While these \emph{learned} query optimizers (LQOs) represent exciting work and have even seen some industry adoption~\cite{steer-qo-at-ms}, we argue that existing LQOs capture at most two of three desirable properties:

\begin{enumerate}
    \item{An LQO should \emph{learn from its mistakes}, quickly and actively incorporating feedback from query plans}.
    \item{An LQO should \emph{work immediately}, without requiring extensive training (i.e., the ``cold start'' problem)}.
    \item{An LQO should \emph{adapt} to changes in schema, workload, and data, without significant human effort}.
\end{enumerate}

For example, the Bao~\cite{marcus2021bao} system learns from its mistakes and adapts to changes, but requires several hours of training data before becoming competitive. The Fastgres system~\cite{fastgres} does not suffer from the cold start problem and can learn from its mistakes, but it cannot adapt to changes in query workload with significant retraining. Three recent systems applying LLMs (or their components) to query optimization~\cite{llm4qo-1, llm4qo-2,akioyamen2024unreasonable} employ fine-tuned LLMs as an oracle for either plan generation or plan selection. However, due to the high cost of fine-tuning an LLM, these approaches cannot immediately learn from their mistakes or continuously improve during runtime.

Why is there no existing system with all the desired properties? Intuitively, LQOs that do not use LLMs tend to have below-par reasoning skills: they must simplify the problem by either requiring large amounts of training data (and therefore having a ``cold start'') or by assuming the schema and workload are fixed (and therefore not being able to adapt). LQOs that do use LLMs can work immediately and adapt to changes, but correcting their mistakes using current techniques is too costly for the critical path of a DBMS.

In this paper, we propose \oursystem, a RAG-based query optimizer that has all three of our desired properties. \oursystem uses an LLM to avoid the cold start problem and adapt to changes. To learn from its mistakes, \oursystem uses a continuously-updated vector database of past query executions that serve as the query optimizer's ``memory''.
{\oursystem} generates a hint-based query plan to guide the query optimizer in a classical database system (e.g., PostgreSQL~\cite{postgres}), where such a hint is optimized and generated using an LLM. {\oursystem} relies on two complementary phases to optimize the hint generation process: \textit{offline LLM fine-tuning} and \textit{online LLM in-context learning}. Since these techniques have higher overhead than traditional query optimizers, SERFQO targets long-running OLAP queries where performance gains can outweigh increased optimization overhead.

To prepare a fine-tuned LLM capable of generating correctly formatted and efficient query plan hints, {\oursystem} adopts a combined supervised fine-tuning (SFT)~\cite{gunel2020supervised} and reinforcement fine-tuning (RFT)~\cite{rft-sft-1} technique. Using SFT, {\oursystem} teaches the LLM the expected hint format. For RFT, {\oursystem} introduces an efficient Query Group Relative Policy Optimization (\qgrpo) approach, based on GRPO~\cite{deepseekmath}, to steer the LLM’s output preferences toward hints that reduce query execution latency. {\qgrpo} defines a latency-based reward model that interacts with the database execution engine to directly measure the corresponding latency of the generated query hint. Besides, {\qgrpo} performs an in-group comparison among multiple hints generated for the same query; it obviates the need for additional labeled datasets as in other RFT algorithms, such as DPO (e.g.~\cite{llm4qo-1}). By not relying on a limited labeled dataset, {\qgrpo}’s fine-tuning encourages LLMs’ self-exploration, which further inspires their reasoning and generalization capabilities to handle previously unseen workloads or shifting database schemas.

Another key innovation of {\oursystem} lies in leveraging online LLM RAG-based in-context learning, in which LLMs adapt to new tasks by conditioning on a few relevant RAG-based examples that are retrieved and provided directly within the input context (i.e., the prompt) without modifying model parameters. 
These examples allow the model to better understand tasks and generate more appropriate outputs~\cite{incontext1,incontext2,incontext3,incontext4}. Importantly, in-context learning is most effective when the retrieved examples closely align with the target query. To facilitate this, {\oursystem} builds and maintains a special vector database of the most relevant query execution records. Additionally, {\oursystem} introduces a self‑evolving feedback paradigm that enables the LLM to learn from historical execution records of the same query. The historical best plan generated by {\oursystem} and its corresponding performance gain over the baseline provide additional contextual information. These components iteratively optimize the prompt, motivating the LLM to generate an optimized hint that minimizes the execution latency.





Our experimental results show that {\oursystem} consistently outperforms both PostgreSQL and two state-of-the-art LQOs, namely Bao~\cite{marcus2021bao} and Balsa~\cite{yang2022balsa}. Our evaluations are based on four workloads: JOB~\cite{leis2015good}, CEB~\cite{flowloss}, Stack~\cite{marcus2021bao}, and TPC-DS~\cite{tpcds}. On CEB, {\oursystem} achieves a 65.05\% reduction in query execution time compared to PostgreSQL, and outperforms LQOs. On Stack, it can reduce up to $93.57\%$ of the execution time for certain queries and $40.10\%$ for overall performance. On TPC-DS, {\oursystem} reduces 13.28\% execution time for all queries. By leveraging a dynamic prompt with a self‑evolving paradigm and successful fine‑tuning, {\oursystem} demonstrates surprising cross‑domain capability on the previously unseen Stack workload, even surpassing LQOs directly trained on that workload in some metric. {\oursystem} also supports flexible workload shifts that current LQOs find challenging.{\oursystem} also supports the cross-DBMS scenario. On MySQL, it achieves a 26.28\% reduction in execution time without requiring extra fine-tuning. Additionally, we observe that {\oursystem} can learn from the provided query plan at the sub-plan level and generate an improved query plan through further reasoning and analysis.
Furthermore, we conduct overhead analysis and ablation studies to evaluate several configurations. Finally, we conduct deeper investigations across various scenarios to test robustness and effects.


We summarize our contributions as follows:
\begin{enumerate}
    \item We are the first to formally define hint-based query optimization with LLMs as a two-phase process: offline LLM fine-tuning and online in-context learning.
    \item We propose an efficient supervised fine-tuning workflow to teach LLMs the hint format using a labeled dataset.
    \item We introduce Query Group Relative Policy Optimization that incorporates a latency-based reward model based on actual query execution and group relative advantage feedback, enhancing LLM reasoning in unseen scenarios.
    \item We present RAG-based prompt optimization, which leverages similar references and historical feedback to dynamically construct prompts and drive continuous self‑evolution of {\oursystem}. To our knowledge, this is the first application of RAG systems to query optimization.
    \item Our experiments demonstrate that {\oursystem} outperforms both LQOs and PostgreSQL and generalizes effectively across multiple workloads and DBMSes without retraining.
\end{enumerate}

\section{Preliminaries}
LLMs are transformer-based models~\cite{transformer} with billions of parameters that generalize well across a wide range of tasks (e.g., question-answering~\cite{questionanswering} and code generation~\cite{codegenerationsurvey}). During the pre-training stage, LLMs learn from a vast corpus of knowledge and capture general understanding capabilities. To address the query optimization task, we exploit the LLM to generate a hint-based query plan, and to adapt the LLM to this specific task, we rely on fine-tuning and RAG-based prompting procedures. First, we provide a brief background on the hint-based query optimization process (Section~\ref{subsec:hint}). Then, we describe how the LLM's generation (inference) process works (Section~\ref{sec:llm-inference}). 
 Finally, we provide an overview of the two main LLM fine-tuning paradigms, namely supervised fine-tuning (Section~\ref{sec:sft}) and reinforcement fine-tuning (Section~\ref{sec:rft}), as well as the RAG retrieval workflow (Section~\ref{subsec:retrival-in-RAG}) we use in our system.

 

\subsection{Hint-based Query Plan Generation}\label{subsec:hint}

Hint-based plan generation is a commonly employed method in many learned query optimizers (LQOs) (e.g.,~\cite{yang2022balsa, zhu2023lero}). For example, LQOs  (e.g.,~\cite{yang2022balsa}) that are integrated with PostgreSQL~\cite{postgresql} employ \textit{pg\_hint\_plan}~\cite{pg_hint_plan} to tweak the PostgreSQL~\cite{postgresql} execution plans by incorporating ``hint'' in SQL comments. The hints support both a join-order mode (simple mode) and a full-plan mode (complete mode). In join-order mode, the hint is represented by a clause beginning with ``\texttt{Leading}'' to specify the join order of relations. For example, the hint: \(\texttt{Leading (a (b c))}\) indicates that \(b \bowtie c\) is performed first, followed by \(a \bowtie (b \bowtie c)\). Then, PostgreSQL will decide the detailed operators plan (e.g., operator type for each join). In full-plan mode, the hint conveys additional information, such as the scan types for tables (e.g., ``\texttt{IndexScan}'', ``\texttt{SeqScan}'') and the join types between tables (e.g., ``\texttt{HashJoin}'', ``\texttt{NestedLoop}''). This hint format is interpreted by PostgreSQL to precisely control query planning and execution. An example of a full-plan mode hint is shown in the following section (see the third part in Figure~\ref{fig:extract_hint}). In our paper, we employ both modes as different query plan generation settings. Once a hint is accepted by the query optimizer, it is turned into a query plan and subsequently used for actual execution.


\subsection{LLM Generation}\label{sec:llm-inference}
Prompts are commonly used to interact with LLMs. A prompt is a piece of text that guides the model in generating more content. It may include a \texttt{System Prompt}, which defines the model’s role or context (e.g., ``You are a database expert performing query optimization task''), and a \texttt{User Prompt}, which specifies particular questions or tasks (e.g., ``Generate a query plan for this query:'' followed by detailed query information). When processing this natural‑language prompt, the tokenizer separates the raw text into words and converts them into a sequence of discrete tokens. Each token is then mapped to an integer according to the model’s vocabulary, after which the LLM can begin processing. The most fundamental input can be represented as a sequence of tokens \(\mathbf{x} = [x_1, \dots, x_n]\).

LLMs often adopt an auto-regressive approach to generate response sequences~\cite{transformer}. Their inference processes are typically based on next-token prediction. Specifically, the original input tokens and the previously generated output tokens from the previous steps are fed into the model to generate a new token at the current step. Assume an input sequence \(\mathbf{x}\) is sent to an LLM parameterized by~\(\theta\). The corresponding output token sequence can be expressed as \(\mathbf{y} = [y_1, \dots, y_T]\), containing \(T\) tokens. This token sequence \(\mathbf{y}\) is then converted back to natural language by the tokenizer. In our context, \(\mathbf{y}\) represents a query hint (See Section~\ref{subsec:hint}). Mathematically, the LLM’s generation process can be expressed by the probability \(p_\theta(\mathbf{y}\mid \mathbf{x})\), which factorizes into the product of conditional probabilities for each token \(y_t\) in the output sequence:
\begin{align}
p_\theta(\mathbf{y} \mid \mathbf{x}) \;=\; p_\theta\bigl(y_1, y_2, \dots, y_T \mid \mathbf{x}\bigr)
\;=\;\prod_{t=1}^T p_\theta\bigl(y_t \,\big|\,
\mathbf{x},\, y_{1:t-1}\bigr)
\label{equ:llm-inference}
\end{align}
For clarity in the rest of the paper, we simplify the token-related notions and hide the tokenizer workflow in the LLM generation process by describing the whole process at the natural-language level. We use the notation \(\mathrm{x}\) instead of \(\mathbf{x}\) to represent the \textit{prompt}, and use \(h\) instead of \(\mathbf{y}\) to directly denote the \textit{generated hint}.


\subsection{Supervised Fine-tuning for LLMs}\label{sec:sft}
Supervised fine-tuning (SFT)~\cite{gunel2020supervised} adjusts the parameters of LLMs by training them on high-quality, labeled data to enhance their ability to perform specific tasks. To improve LLM performance in our hint-based query plan generation task, a training set \(\mathbb{D}_{\mathrm{sft}}\) comprising pairs of prompt \(\mathrm{x}\) and corresponding pre-generated hint output \(h^\ast\) needs to be constructed (We describe the details in Section~\ref{sec:Supervised Fine-tuning for LLMs}). Here, \(h^\ast\) denotes the reference hint, distinguishing it from the LLM-generated hint \(h\). During supervised fine-tuning, we maximize the conditional predictive probability \(p_{\theta}(h^\ast \mid \mathrm{x})\) by minimizing the expectation \(\mathbb{E}\) of the negative log-likelihood:
\begin{align}
\mathcal{L}_{\mathrm{sft}}(\theta)
= \mathbb{E}_{(\mathrm{x},h^\ast)\sim \mathbb{D}_{\mathrm{sft}}}
\bigl[-\log p_\theta(h^\ast \mid \mathrm{x})\bigr] \label{equ:sft}
\end{align}
Using SFT, the LLM is expected to become familiar with our task, i.e., learn the format of the query hint output and internal logic for generating an effective query plan.

\subsection{Reinforcement Fine-tuning for LLMs}\label{sec:rft}

Reinforcement fine-tuning (RFT) provides an alternative method for enhancing the specific capabilities of LLMs by employing reinforcement learning to update their parameters. RFT has been proven effective in further improving the reasoning abilities of LLMs after the SFT stage (e.g.,~\cite{rft-sft-1}). In our task, reasoning ability reflects an improved generalization capability when \textit{handling previously unseen queries}. This implies that, in cross-domain scenarios (e.g., switching between completely different query workloads, such as Stack~\cite{marcus2021bao} and CEB~\cite{flowloss}), LLMs with RFT are expected to generate hints indicating better query plans than those without RFT. Several RFT algorithms have been developed, such as Proximal Policy Optimization (PPO)~\cite{schulman2017proximal}, Direct Preference Optimization (DPO)~\cite{rafailov2023direct}, and the recently proposed Group Relative Policy Optimization (GRPO)~\cite{deepseekmath}. During fine-tuning, PPO relies on 
a trained value model to objectively determine whether the generated output is \texttt{preferred}  or \texttt{non\_preferred}. This value model introduces additional memory requirements and computational overhead~\cite{deepseekmath}. While DPO eliminates the need for a separate value model to reduce complexity, it requires paired preference data (consisting of \texttt{preferred} and \texttt{non\_preferred} output) to guide the LLM's optimization process. 
Constructing such a preference-pair dataset before fine-tuning introduces an additional workload for data labeling, which is complex for our query optimization task. Moreover, although DPO increases the relative probability of preferred completions compared to dispreferred ones, it may reduce the model's absolute likelihood of generating preferred completions~\cite{dpo-drawback}.

To avoid these limitations, we rely on GRPO~\cite{deepseekmath}, a variant of PPO that enhances computational efficiency and robustness. GRPO incorporates group-based comparisons among generated responses to eliminate the need for an explicit value model, as required by PPO, and also removes the need for labeled data, as in DPO. GRPO generates multiple outputs from the same prompt, computes rewards for these outputs based on the user-defined reward function, and then evaluates their in-group advantages~$\hat{A}$. Specifically, the GRPO algorithm first fixes the current policy model\footnote{RFT treats the prompt as the environment state and each token generation as an action. So we use the reinforcement learning policy model \(\pi_\theta\) notation rather than \(p_\theta\).} as \(\pi_{\theta_{\text{old}}}\). This naming distinguishes it from the policy model to be updated, \(\pi_{\theta}\). Then, for each prompt \(\mathrm{x}\) in the GRPO training dataset \(\mathbb{D}_{\mathrm{GRPO}}\), an LLM generates a group of \(G\)  hint plans \(\{h_1, h_2, \dots, h_G\}\) based on \(\pi_{\theta_{\text{old}}}\). Then, these hints are executed with a database engine, and their corresponding execution latencies are used to optimize the policy model \(\pi_{\theta}\) by maximizing the following objective:
\begin{equation}\label{equ:grpo-target}
\begin{aligned}
\mathcal{J}_{\text{GRPO}}&(\theta) \approx \mathbb{E}_{\mathrm{x} \sim \mathbb{D}_{\mathrm{GRPO}}, \{h_i\} \sim \pi_{\theta_{\text{old}}}} \\
&\quad \frac{1}{G}\sum_{i=1}^G \Biggl\{
f\!\left(\frac{\pi_\theta(h_i \mid \mathrm{x})}{\pi_{\theta_{\text{old}}}(h_i \mid \mathrm{x})},\,\epsilon\right) \hat{A}_i 
- \beta\,\mathbb{D}_{\text{KL}}[\pi_\theta\|\pi_{\text{ref}}]
\Biggr\}
\end{aligned}
\end{equation}
where function \(f\) and hyperparameter \(\epsilon\) are employed to stabilize fine-tuning~\cite{schulman2017proximal}. To prevent the updated policy \(\pi_\theta\) from excessively deviating from the initial policy denoted as reference policy \(\pi_{\text{ref}}\), a Kullback–Leibler (KL) divergence regularization~\cite{kullback1951information,deepseekmath} term weighted by the hyperparameter \(\beta\) is applied.


\subsection{Retrieval via Embedding Similarity}\label{subsec:retrival-in-RAG}
Retrieval-augmented generation (RAG)~\cite{taipalus2024vectorvectordbsurvey} enhances large language models (LLMs) by incorporating external knowledge retrieved from a vector database~\cite{wang2021milvus}, thereby supplementing the prompt with relevant information without altering the model's weights. This approach effectively mitigates the limitations imposed by the static knowledge of standalone LLMs. Consequently, the selection of the most relevant content from the vector database is critical to performance. To facilitate finer-grained control, the retrieved content can be segmented into smaller, semantically meaningful snippets when necessary~\cite{mao2024fitragsplit,csakar2025maximizingragsplit}. In our context, each snippet corresponds to a reference query $d_i$, and we define the full set of reference queries as follows:
\(
\mathcal{D} = \{ d_1, d_2, \dots, d_N \}.
\)
For each reference query \(d_i\), we use an encoder \(\phi(\cdot)\) to map it into a \(m\)-dimensional embedding space as:
\(
\mathbf{e}_i = \phi(d_i) \in \mathbb{R}^m.
\)
All embeddings can be organized into a matrix:
\(
\mathbf{E} = \left[\mathbf{e}_1, \mathbf{e}_2, \dots, \mathbf{e}_N\right] \in \mathbb{R}^{m \times N}.
\)
Given an query \(\mathbf{u}\) for vector database, its embedding is computed as:
\(
\mathbf{e}_\mathbf{u} = \phi(\mathbf{u}) \in \mathbb{R}^m.
\)
The similarity between $\mathbf{u}$ and each $d_i$ can be computed using cosine similarity as follows:
\begin{align}
\operatorname{sim}(\mathbf{u}, d_i) = \frac{\mathbf{e}_\mathbf{u}^\top \mathbf{e}_i}{\|\mathbf{e}_\mathbf{u}\| \, \|\mathbf{e}_i\|}.
\end{align}
Based on these similarity scores, we rank and pick the top \(k\) references to form the retrieved set \(\mathcal{R}\), where:
\(
\mathcal{R} \subset \mathcal{D}, |\mathcal{R}| = k.
\)

\section{System Overview}
In this section, we first formalize the definition of optimizing queries using {\oursystem} (Section~\ref{sec:llmqo definition}). Then, we give an overview of the {\oursystem}'s workflow and its key components (Section~\ref{sec:system framework}).

\subsection{Problem Statement}
\label{sec:llmqo definition}
Our ultimate goal is to reduce query execution latency using hints generated by LLMs. We achieve this through two complementary phases: \textit{offline fine-tuning for the LLM abilities to generate hints}, and \textit{online in-context learning for the LLM to generate optimal hints}.

\noindent\textbf{Definition 1: Offline LLM Fine-tuning for Hints Generation.}  
In this phase, we aim at fine-tuning an LLM, parameterized by $\theta_{\text{LLM}}$, such that, for any input SQL query \(Q \in \mathcal{Q}\) and the relevant database statistics \(\texttt{stats}\) that help plan the execution of this query (e.g., table cardinalities), the LLM is able to generate a hint $h$ that optimizes the execution of $Q$, with the objectives of (1) ensuring correct hint syntax (valid to become a plan) and (2) steering the LLM's preference towards generating hints that minimize query execution latency.
These objectives can be formally defined as:
\begin{align}
  \arg\max_{\theta_{\text{LLM}}}\;&
    \mathbb{E}_{\mathcal{Q},\mathrm{stats}}\bigl[\log p_{\mathrm{correct}}(h)\bigr]
    &&\text{(Syntax Correctness)}
    \label{equ:syntax-correctness}\\
  \arg\min_{\theta_{\text{LLM}}}\;&
    \mathbb{E}_{\mathcal{Q},\mathrm{stats}}\bigl[\mathrm{Latency}(Q,h)\bigr]
    &&\text{(Latency Preference)}
    \label{equ:latency-preference}
\end{align}
where \(\mathbb{E}_{\mathcal{Q},\mathrm{stats}}\) denotes the expectation over the joint data distribution of input SQL queries \(\mathcal{Q}\) and database statistics \(\texttt{stats}\). This offline fine-tuning phase is achieved using SFT (see Section~\ref{sec:sft}) and GRPO (see Section~\ref{sec:rft}) as shown later. 


\noindent\textbf{Definition 2: Online LLM In-context Learning for Hints Generation.}  
In this phase, for any input SQL query \(Q \in \mathcal{Q}\) and its relevant database statistics \(\texttt{stats}\) mentioned in Definition~1, we aim at constructing an optimized prompt~$\mathrm{x}^\ast$ (noted as \(P\) in this section), during the runtime, that guides the fixed-parameter LLM to generate a hint $h$ that minimizes the execution latency of this query (i.e., \(h\gets\text{LLM}_{\theta_{\text{LLM}}}(P)\)). Formally, this corresponds to solving the following optimization problem:
\begin{align}\label{eq:online-latency}
  \arg\min_{P}\;&
    \mathrm{Latency}(Q,h) &&\text{(Prompt Optimization)}
\end{align}
The optimized prompt $P$ for any query is constructed based on the relevant database statistics \texttt{stats} of this query and a few relevant query references \(\mathcal{K}\) that are dynamically retrieved from an up-to-date RAG vector database (see Section~\ref{subsec:retrival-in-RAG}), as shown later.

\subsection{Overview of {\oursystem}}
\label{sec:system framework}

\begin{figure}[tb]
    \centering
\includegraphics[width=1\linewidth]{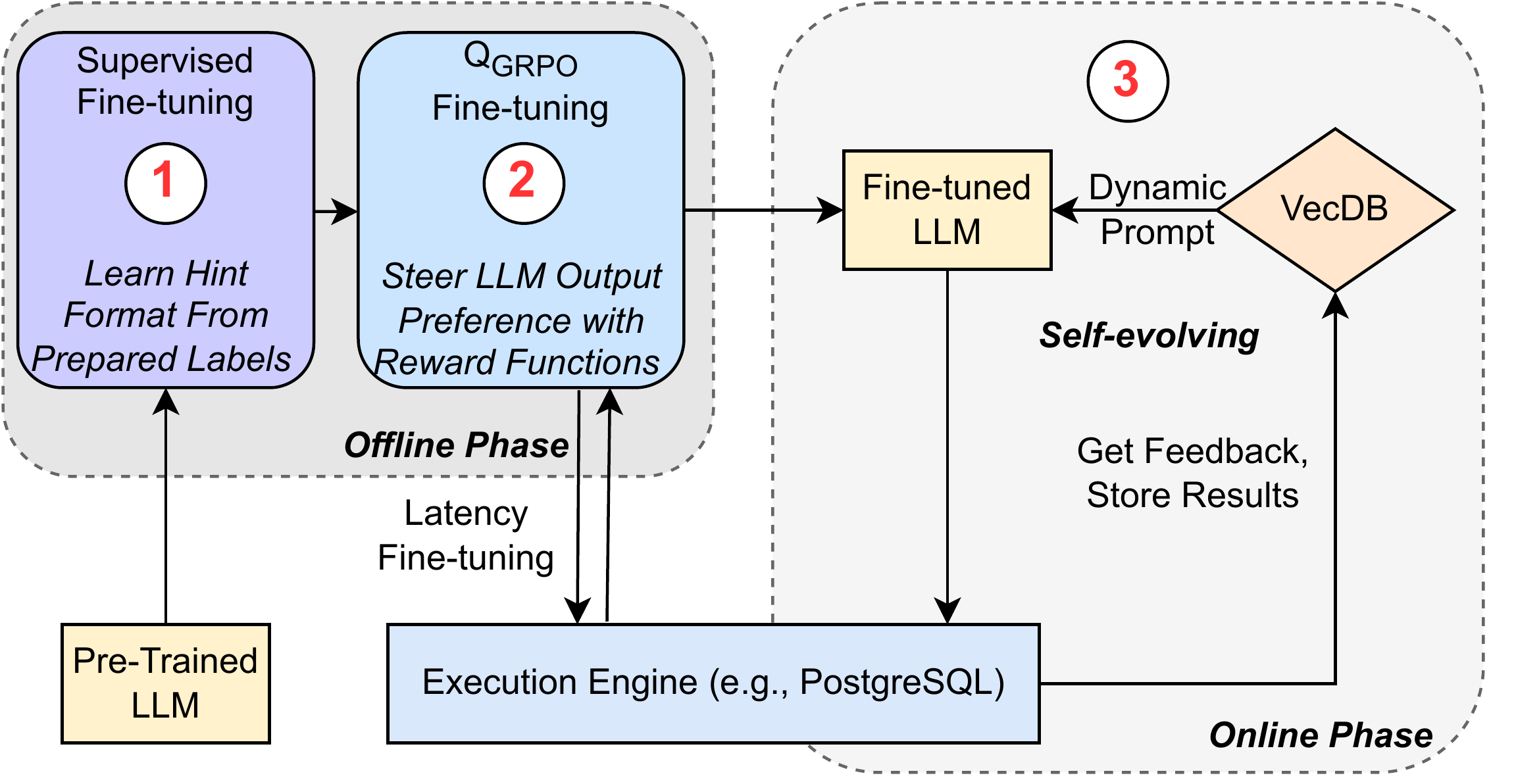}
    \caption{System overview of {\oursystem}.}
    \label{fig:system_overview}
\end{figure}

Figure~\ref{fig:system_overview} shows the overview of {\oursystem}, which comprises three stages, each mapping to an optimization problem in Section~\ref{sec:llmqo definition}. Corresponding to Equation~\ref{equ:syntax-correctness}, the first stage is \textbf{Supervised Fine-Tuning (SFT)} (see Section~\ref{sec:Supervised Fine-tuning for LLMs}). SFT teaches the model the correct format of query hints by training LLMs on a pre-constructed dataset containing pairs of input prompts (user prompt as in Section~\ref{sec:llm-inference}) and output hints (training set \(\mathbb{D}_{\mathrm{sft}}\) in Section~\ref{sec:sft}). SFT lays the foundation of our system because any further improvements depend on LLMs’ ability to generate syntactically correct hints that a database execution engine (e.g., PostgreSQL) can accept. Corresponding to Equation~\ref{equ:latency-preference}, the second stage is \textbf{Query Group Relative Policy Optimization ({\qgrpo}) fine-tuning} (see Section~\ref{sec:qgrpo}). In {\qgrpo}, we incorporate latency feedback from actual query executions using the database engine (training dataset \(\mathbb{D}_{\mathrm{GRPO}}\) in Section~\ref{sec:rft}). By designing a latency-based reward model and employing group relative advantage feedback, we steer LLMs’ preferences toward generating hints with lower latency. As a reinforcement-learning process in a dynamic environment, {\qgrpo} enhances reasoning ability for unseen workloads. These first two stages occur during the offline phase. Corresponding to Equation~\ref{eq:online-latency}, the third stage is a \textbf{RAG-based Prompt Optimization} online phase, where the output prompt will be used to generate the final hint used for the query optimization. In this phase, we maintain a continuously updated RAG system incorporating live execution records (see Section~\ref{sec:RAG-SYSTEM_DESIGN}). For each query \(Q\), we dynamically construct a prompt containing execution records of similar queries as references and the historical execution analysis (see Section~\ref{sec:Self-evolving Feedback Paradigm}). Upon repeated encounters with \(Q\), the prompt is continuously optimized. This self‑evolving procedure enables {\oursystem} to further minimize the execution latency.

\section{SFT for Learning Hint Format}
\label{sec:Supervised Fine-tuning for LLMs}

In this section, we primarily focus on employing supervised fine-tuning to enable LLMs to understand the query optimization task and to learn the expected hint format. 

\begin{figure*}[tb]
    \centering
    \includegraphics[width=1\linewidth]{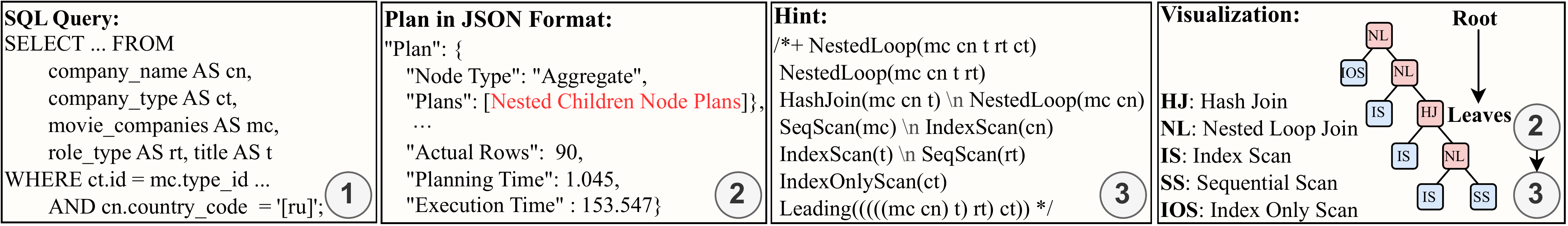}
    \caption{Given the SQL text and other information (Step 1), we extract the query hint from the query plan in the JSON format (Step 2) and simplify the hint structure (Step 3). The rightmost sub-figure shows how we simplify the plan representation.}
    \label{fig:extract_hint}
\end{figure*}

As mentioned in Section~\ref{sec:sft}, we need to construct a training set \(\mathbb{D}_{\mathrm{sft}}\) that consists of pairs of prompt \(\mathrm{x}\) and corresponding generated hint output \(h^\ast\) (i.e., labeled data). We define the \(\mathbb{D}_{\mathrm{sft}}\) collection process in Algorithm~\ref{algorithm:sft-dataset-prep}. It shows the sequence of operations we perform for each query \(Q_i \in \mathcal{Q}_{train}\) to build the training set. Also, Figure~\ref{fig:extract_hint} shows a running example of the intermediate outputs of these different operations. First, we execute the query \(Q_i\) using the database execution engine to retrieve the detailed query execution plan and its intermediate statistics \(\mathcal{E}_i\) (line~2) (using commands like EXPLAIN ANALYZE in PostgreSQL). We return the plan in a structured and nested JSON format to easily parse its details. Figure~\ref{fig:extract_hint} (Part~2) shows an example of the JSON output of only some statistics of the ``Aggregate'' operator in the example query (Note that the remaining statistics for the Aggregate operator and other operators are summarized in red as ``Nested Children Node Plans'' for clarity). Then, we extract the tables \(\mathcal{T}_i\) used in \(Q_i\) and get their cardinalities \(\mathcal{C}_i\) from the JSON output (lines 3 and 4). Next, the original query \(Q_i\) is concatenated with the extracted cardinalities \(\mathcal{C}_i\) to construct the prompt \(\mathrm{x}_{Q_i}\) (line 5). Finally, we obtain the generated hint \(h^\ast\) (i.e., labeled output) for the query \(Q_i\) (line 6), pair it with the constructed prompt \(\mathrm{x}_{Q_i}\), and add them to the training set \(\mathbb{D}_{\mathrm{sft}}\). Figure~\ref{fig:extract_hint} (Part~3) shows an example of the constructed hint for the plan tree structure highlighted in the rightmost part of the figure.
After obtaining \(\mathbb{D}_{\mathrm{sft}} = \{(\mathrm{x}_{Q_i}, h^\ast_i)\}\), we maximize the conditional predictive probability \(p_{\theta}(h^\ast|\mathrm{x})\) of the LLM as described in Section~\ref{sec:sft}.

\begin{algorithm}[h]
\caption{Supervised Fine-tuning Dataset Preparation}
\label{algorithm:sft-dataset-prep}
\begin{algorithmic}[1]
\Statex \hspace*{-\algorithmicindent}\textbf{Input:}  Database engine $\mathcal{DB}$, Training query set $\mathcal{Q}_{train} = \{Q_1, Q_2, \dots, Q_n\}$
\Statex \hspace*{-\algorithmicindent} \textbf{Output:} SFT training dataset $\mathbb{D}_{\mathrm{sft}} = \{(\mathrm{x}_{Q_i}, h^\ast_i)\}$
\For {each $Q_i \in \mathcal{Q}_{train}$}
    \State $\mathcal{E}_i \gets \mathcal{DB}(Q_i)$ \Comment{Generate execution record for $Q_i$}
    \State $\mathcal{T}_i \gets \text{ExtractTables}(\mathcal{E}_i)$ \Comment{Extract used tables}
    
    \State $\mathcal{C}_i \gets \text{GetCardinalities}(\mathcal{T}_i, \mathcal{E}_i)$ \Comment{Retrieve table cardinalities}
    \State $\mathrm{x}_{Q_i} \gets \text{ConstructPrompt}_{\mathrm{sft}}( Q_i, \mathcal{C}_i)$ \Comment{Construct SFT prompt}
    \State $h^\ast_i \gets \text{ExtractHint}(\mathcal{E}_i)$ \Comment{Extract hint as label}
    \State Append $(x_i, h^\ast_i)$ to $\mathbb{D}_{\mathrm{sft}}$
\EndFor
\State \textbf{Return} $\mathbb{D}_{\mathrm{sft}}$
\end{algorithmic}
\end{algorithm}

\section{Query Group Relative Policy Optimization}\label{sec:qgrpo}

\begin{figure}[t]
    \centering
    \includegraphics[width=1\linewidth]{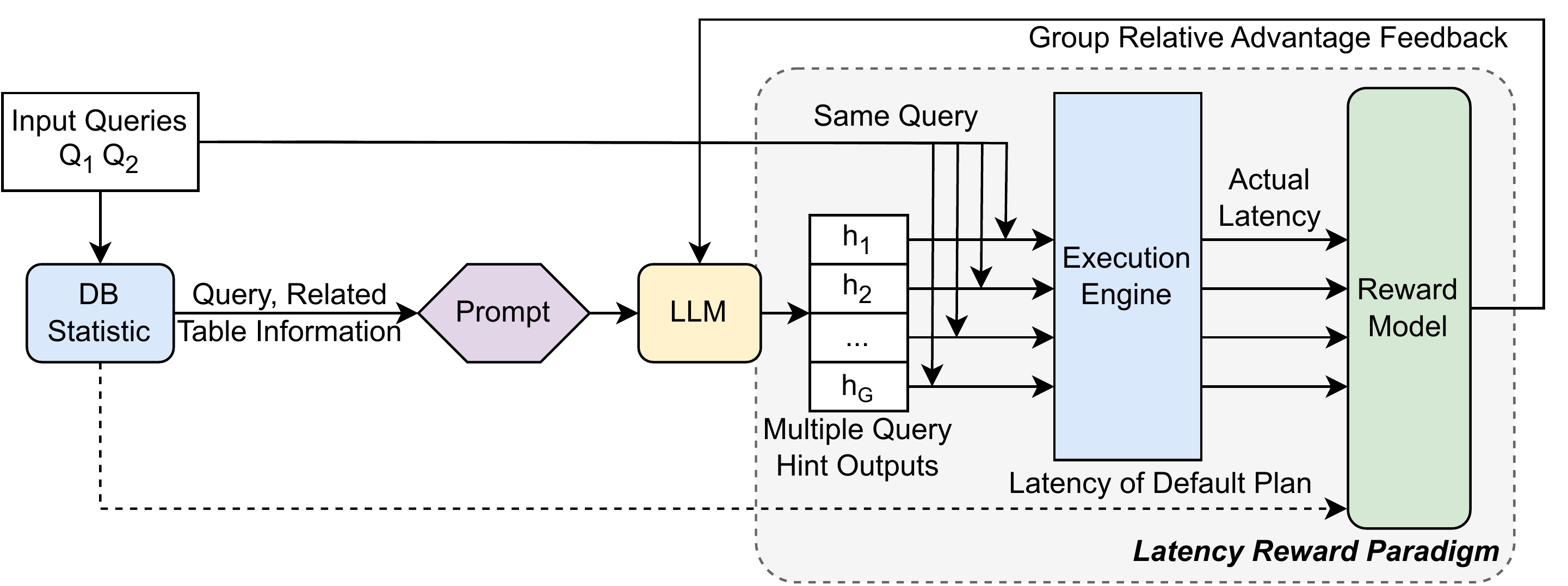}
    \caption{Overview of the {\qgrpo} workflow with Latency-based Reward Paradigm.}
    \label{fig:qgrpo-framework}
\end{figure}

This optimization stage further enhances the LLM's ability to perform query optimization tasks by steering the model's output toward a query hint with reduced latency. We propose Query Group Relative Policy Optimization ({\qgrpo}) based on GRPO~\cite{deepseekmath} to use reinforcement learning to continuously fine-tune \(\theta_{\mathrm{sft}}\), which was obtained from the previous supervised fine-tuning stage. The key intuition is to have the LLM generate multiple hint outputs \(\{h_1, h_2, \dots, h_G\}\) for a given prompt \(\mathrm{x}\). Then, we use gradient descent to update the model's parameters by computing the in-group advantages of these hint outputs, which correspond to relative latency improvement rewards in our context. As a result, the policy model is optimized to an updated parameter set \(\theta_{\mathrm{GRPO}}\). Figure~\ref{fig:qgrpo-framework} illustrates the {\qgrpo} workflow and the latency-based reward paradigm.

\subsection{Latency-based Reward Model}\label{subsec:latency-based-reward-model}
For a given query, the ultimate objective of query optimization is to reduce the latency of the query plan. To directly leverage feedback on query execution latency, we propose a latency-based reward model derived from real query execution and integrated with reinforcement fine-tuning for LLMs.

\begin{algorithm}[t]
\caption{Latency-based Reward Model}
\label{alg:query_hint_reward}
\begin{algorithmic}[1]
\Statex \hspace*{-\algorithmicindent}\textbf{Input:} Query $Q$, corresponding multiple LLM outputs as hint plans$\{h_1, h_2, \dots, h_G\}$, latency reward function $R_\text{Latency}$ 
\Statex \hspace*{-\algorithmicindent}\textbf{Output:} Rewards $\{r_1, r_2, \dots, r_G\}$
\State Baseline latency: $t_d\gets \text{ExecEngine}(Q)$\Comment{without hint}
\For{each Hint Output $i \in \{1,2,\dots,G\}$}
    \State Actual latency: $t_i \gets \text{ExecEngine}(Q, h_i)$\Comment{with hint}
    \State latency improvement ratio: $\triangle_{t_i} \gets \frac{t_d}{t_i}$\Comment{Improvement rate}
    \State Latency reward: $r_i \gets R_\text{Latency}(\triangle_{t_i})$ 
\EndFor
\State \textbf{Return:} $\{r_1, r_2, \dots, r_G\}$
\end{algorithmic}
\end{algorithm}

Algorithm~\ref{alg:query_hint_reward} illustrates the latency-based reward model used in {\qgrpo}. We compute the reward based on the actual latency speedup relative to the baseline latency obtained by executing the same query without any query hint. First, we execute the query without applying a query hint on the database execution engine to obtain the baseline latency \(t_d\) (line 1). Each LLM-generated hint output \(h_i\) is concatenated with the query \(Q\) and executed to record the actual latency \(t_i\) (line 3). We then calculate the latency improvement ratio \(\triangle_{t_i}\) as the ratio between the baseline latency \(t_d\) and the actual latency \(t_i\) (line 4). A larger value of \(\triangle_{t_i}\) indicates a better query hint. Finally, the latency reward \(r_i\) is derived using the designed latency reward function \(R_{\text{Latency}}\) (line 5), which is defined as follows:
\begin{align}
R_{\text{Latency}}(\triangle_{t_i}) = \tanh\left(\ln(\triangle_{t_i})\right)
\label{equ:reward-function-definition}
\end{align}
where the logarithmic function \(\ln\) is introduced to achieve symmetry reward. For example, a two-fold improvement (\(\Delta_{t_i} = 2\)) yields \(\ln(2) \approx 0.693\), while a two-fold degradation (\(\Delta_{t_i} = 0.5\)) results in \(\ln(0.5) \approx -0.693\). Although the logarithmic function dampens the effect of huge differences—since some excellent plans or poor plans can lead to order-of-magnitude speedups or slowdowns—a hyperbolic tangent function \(\tanh\) is further employed to restrict the reward range to the interval \([-1, 1]\). Even when the subsequent procedure computes advantages based on in-group comparisons along with the normalization in Equation~\ref{equ:advanteage}, reducing the sensitivity of the reward to extreme deviations can improve the overall stability of the fine-tuning procedure.

Suppose the current latency \(t_i\) is smaller than the baseline latency \(t_d\). In that case, the latency improvement ratio is larger than 1 (\(\triangle_{t_i} > 1\)), leading to a positive reward in the end for this LLM-generated output, and vice versa. These rewards guide the LLM's preferences toward generating hints that reduce query latency.

\subsection{Group Relative Advantage Feedback}
Another key process in {\qgrpo} is the group relative advantage feedback. Based on the previously calculated rewards \(\{r_1, r_2, \dots, r_G\}\), we compute the advantage \(A_i\) for each hint response \(h_i\), reflecting how much better or worse it is compared to others in the group:
\begin{equation}
    A_i = \frac{r_i - \text{mean}(\{r_1, r_2, \dots, r_G\})}{\text{std}(\{r_1, r_2, \dots, r_G\})}\label{equ:advanteage}
\end{equation}
Here, \(\text{mean}(\{r_1, r_2, \dots, r_G\})\) denotes the average reward of the group, and \(\text{std}(\{r_1, r_2, \dots, r_G\})\) denotes the standard deviation of rewards within the group. Subsequently, we compute the ratio of the probability of generating the response \(h_i\) under the new policy \(\pi_{\theta}\) to that under the old policy \(\pi_{\theta_{\text{old}}}\), denoted as \(\frac{\pi_{\theta}(h_i \mid \mathrm{x})}{\pi_{\theta_{\text{old}}}(h_i \mid \mathrm{x})}\). This ratio indicates the degree to which the new policy diverges from the old policy for a given response, thereby guiding the direction of policy updates.  {\qgrpo} employs gradient descent to iteratively maximize the optimization objective defined in Equation~\ref{equ:grpo-target}. The simplified gradient\footnote{This simplified gradient assumes \(\pi_{\theta_{\text{old}}} = \pi_{\theta}\), as in~\cite{deepseekmath}.} used for updating model can be expressed as:
\begin{equation}
    GC_{\mathrm{GRPO}}(\mathrm{x}, h, \pi_{\theta}, \pi_{\text{ref}}) = \hat{A}_{i} + \beta\left(\frac{\pi_{\text{ref}}(h_i \mid \mathrm{x})}{\pi_{\theta}(h_i \mid \mathrm{x})} - 1\right)
    \label{equ:gradient-grpo}
\end{equation}

Algorithm~\ref{alg:grpo} presents the complete fine-tuning process of {\qgrpo}. This algorithm first initializes the policy model \(\pi_{\theta}\) and the reference policy model \(\pi_{\text{ref}}\) as \(\pi_{\text{sft}}\) (line 1), and iterates over batches to update \(\pi_{\theta}\) (lines 2–9). Specifically, we first save the current policy model as \(\pi_{\theta_{\text{old}}}\) to distinguish it from an updating model \(\pi_{\theta}\) (line 3). Then, the LLM generates \(G\) query hints as outputs \(\{h_i\}_{i=1}^{G} \sim \pi_{\theta_{\text{old}}}\) sampling from the saved policy model \(\pi_{\theta_{\text{old}}}\) (line 6). Next, we compute the rewards \(\{r_i\}_{i=1}^{G}\) for each generated hint using the latency-based reward model described in Section~\ref{subsec:latency-based-reward-model} (line 7). Then, we calculate the group relative advantage \(\{A_i\}_{i=1}^{G}\) for each generation according to Equation~\ref{equ:advanteage} (line 8). After computing these values for the batch, we update the LLM parameters via gradient descent with \(\pi_{\text{ref}}\) and \(\pi_{\theta}\) by using Equation~\ref{equ:gradient-grpo} (line 9).

\begin{algorithm}[h]
\caption{Query Group Relative Policy Optimization}
\label{alg:grpo}
\begin{algorithmic}[1]
\Statex \hspace*{-\algorithmicindent}\textbf{Input:} Initial policy model $\pi_{\theta_{\mathrm{sft}}}$, reward function $r_{\phi}$, prompts $\mathcal{P}$
\Statex \hspace*{-\algorithmicindent}\textbf{Output:} Optimized policy model $\pi_{\theta_{\mathrm{GRPO}}}$
\State $\pi_{\theta} \gets \pi_{\theta_{\mathrm{sft}}}$, $\pi_{\text{ref}} \gets \pi_{\theta_{\mathrm{sft}}}$ \Comment{Initialization}
\For {step = $1, \dots, M$}
    \State Save current policy model by $\pi_{\theta_\text{old}} \gets \pi_\theta$
    \State Sample a batch $\mathcal{P}_b$ from $\mathcal{P}$
    \For{(query, table information) pair $\mathrm{x} = (Q, \mathcal{T}) \in \mathcal{P}_b$}
        \State Sample G generated hints $\{h_i\}_{i=1}^{G} \sim \pi_{\theta_{\text{old}}}(\cdot \mid \mathrm{x})$
        \State Compute rewards $\{r_i\}_{i=1}^{G}$ through $r_i = r_{\phi}(h_i)$
        \State Compute relative advantage $\{A_i\}_{i=1}^{G}$ via Equation~\ref{equ:advanteage}
    \EndFor
    \State Update $\pi_\theta$ by maximizing the GRPO objective \Comment{via gradient descent in Equation~\ref{equ:gradient-grpo}, using $\pi_{\text{ref}}$}
\EndFor

\State \textbf{Return:} $\pi_\theta$ as $\pi_{\theta_{\mathrm{GRPO}}}$
\end{algorithmic}
\end{algorithm}
\section{RAG-based Prompt Optimization}\label{sec:RAG-SYSTEM_DESIGN}\label{sec:Overview of LERAG}
After the previous offline fine-tuning procedures, we expect our system to continuously evolve during the online stage to generate query hints, which lead to lower query execution latency. For a fixed fine-tuned LLM, different input prompts for the same query $Q$ can result in varying qualities of hint outputs, which in turn influence the execution latency of the resulting query plans. As formalized in Equation~\ref{eq:online-latency}, by improving the prompt $\mathbf{x}$ to an optimized prompt $\mathbf{x}^\ast$, the LLM generates a query hint $h$ that minimizes the execution latency for $Q$. In this section, we present a RAG-based prompt optimization method. Our RAG workflow dynamically constructs the prompt $\mathbf{x}$ for the LLM based on historical execution records of previously generated query plans. This self‑evolving paradigm enables online in-context learning for hint generation.


\begin{figure}[tb]
    \centering
    \includegraphics[width=1\linewidth]{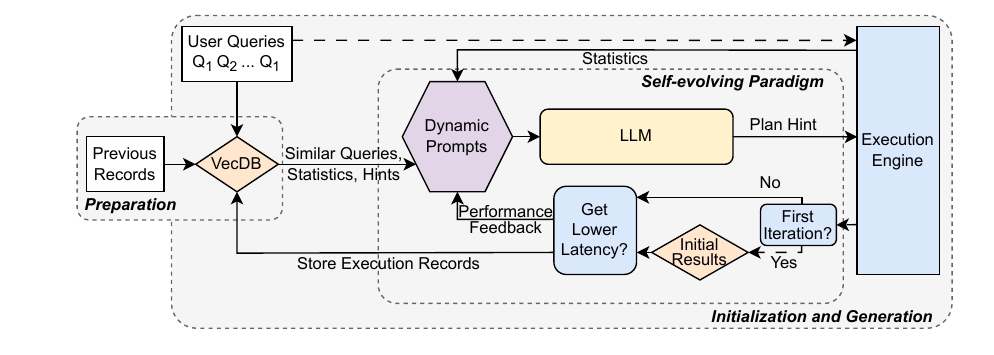}
    \caption{Overview of {\oursystem}'s RAG component.}
    \label{fig:rag-framework}
\end{figure}

\subsection{Architecture of Self-evolving RAG}\label{subsec:Architecture}
Figure~\ref{fig:rag-framework} illustrates the complete pipeline of {\oursystem}’s RAG-based prompt optimization (the self‑evolving phase in Section~\ref{sec:system framework}). It comprises three stages: \textit{Preparation}, \textit{Initialization}, and \textit{Generation}.

During the \textit{Preparation} stage, we collect previous query execution records (e.g., execution times and corresponding query plans generated by PostgreSQL’s default optimizer). Notably, {\oursystem} does not require a predefined workload for this stage. All collected records are loaded into the vector database (VecDB) to bootstrap the system. These execution records serve as reference examples that can be incorporated into the prompt \(\mathrm{x}\) to enhance LLM generation.

After the \textit{Preparation} stage, when a query \(Q\) is encountered for the first time by our RAG system, it triggers the \textit{Initialization} stage (denoted by the dashed line). In this stage, \(Q\) is executed directly on the database engine (e.g., PostgreSQL without applying any query hint \(h\)), and the corresponding execution record is stored in VecDB.


For subsequent encounters with \(Q\), our RAG system enters the \textit{Generation} stage. Here, \(Q\) is sent to the database engine to extract necessary statistics \(stats\) (e.g., table cardinalities and filter selectivities) without actual execution. These statistics are then incorporated into the prompt \(\mathbf{x}\) to provide additional context for better planning. In parallel, \(Q\) is sent to VecDB to retrieve \(k\) reference queries via similarity search (as in Section~\ref{subsec:retrival-in-RAG}), along with their execution records. These reference records are also used to dynamically construct the prompt \(\mathbf{x}\) (see Section~\ref{sec:Prompts Design}). Next, the LLM generates a query hint \(h\) based on this prompt. The query \(Q\), together with the generated hint \(h\), is executed on the database engine. If the new execution record achieves lower latency than the historical record for \(Q\), the latest record replaces the old one in VecDB. If our RAG system encounters the same query \(Q\) again, the stored current best history record for \(Q\) is retrieved from VecDB and used to compute feedback information (e.g., performance gain \(\eta\) over PostgreSQL’s default plan for \(Q\)), which is then incorporated into the dynamic prompt. This feedback paradigm drives continuous self‑evolution, iteratively improving future plan generation (see Section~\ref{sec:Self-evolving Feedback Paradigm}).

\subsection{Vector Database Storage Design}
\label{sec:RAG Storage Design}

\begin{figure}[tb]
    \centering  \includegraphics[width=\linewidth]{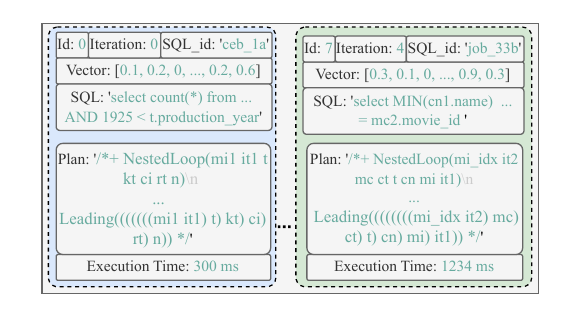}
    \caption{A visual example of storing execution records.}
    \label{fig:collection_example}
\end{figure}

Figure~\ref{fig:collection_example} shows how we store execution records in VecDB. For each record, we maintain seven properties: \texttt{Id}, \texttt{Iteration}, \texttt{Vector}, \texttt{SQL\_id}, \texttt{SQL}, \texttt{Plan}, and \texttt{Execution Time}. \texttt{Id} serves as the primary key. \texttt{Iteration} denotes the current iteration of the online pipeline; for the \textit{Preparation} and \textit{Initialization} stages, it is set to 0 to distinguish them from the \textit{Generation} stage, whose iterations begin at 1. \texttt{Vector} contains the embedding of the original \texttt{SQL} used for similarity search. \texttt{SQL\_id} is a unique identifier for each query, comprising the benchmark name and the internal query name (e.g., \texttt{job\_33b}). It is used to distinguish the current query from its reference queries during similarity search, ensuring that only similar yet distinct queries are retrieved from the vector database.
 \texttt{SQL} denotes the raw query text. \texttt{Plan} represents the query plan extracted from the execution record using the workflow in Figure~\ref{fig:extract_hint}. Finally, \texttt{Execution Time} stores the actual execution latency.

When a new record arrives, our VecDB inserts it according to the layout described above. To extract relevant information, the input query \(Q\) is first encoded as a fixed-length vector using an encoder model (e.g., \textit{SentenceTransformer}~\cite{han2021transformer}). Then, VecDB retrieves the \(k\) most related queries \(\{Q_1, Q_2, \ldots, Q_k\}\) (or fewer if only fewer similar queries exist) based on cosine similarity of their embeddings. Finally, these queries and their associated execution records are used to construct the dynamic prompt \(\mathbf{x}\).


\subsection{Dynamic Prompt Construction}
\label{sec:Prompts Design}

\begin{figure}[tb]
    \centering
    \includegraphics[width=0.8\linewidth]{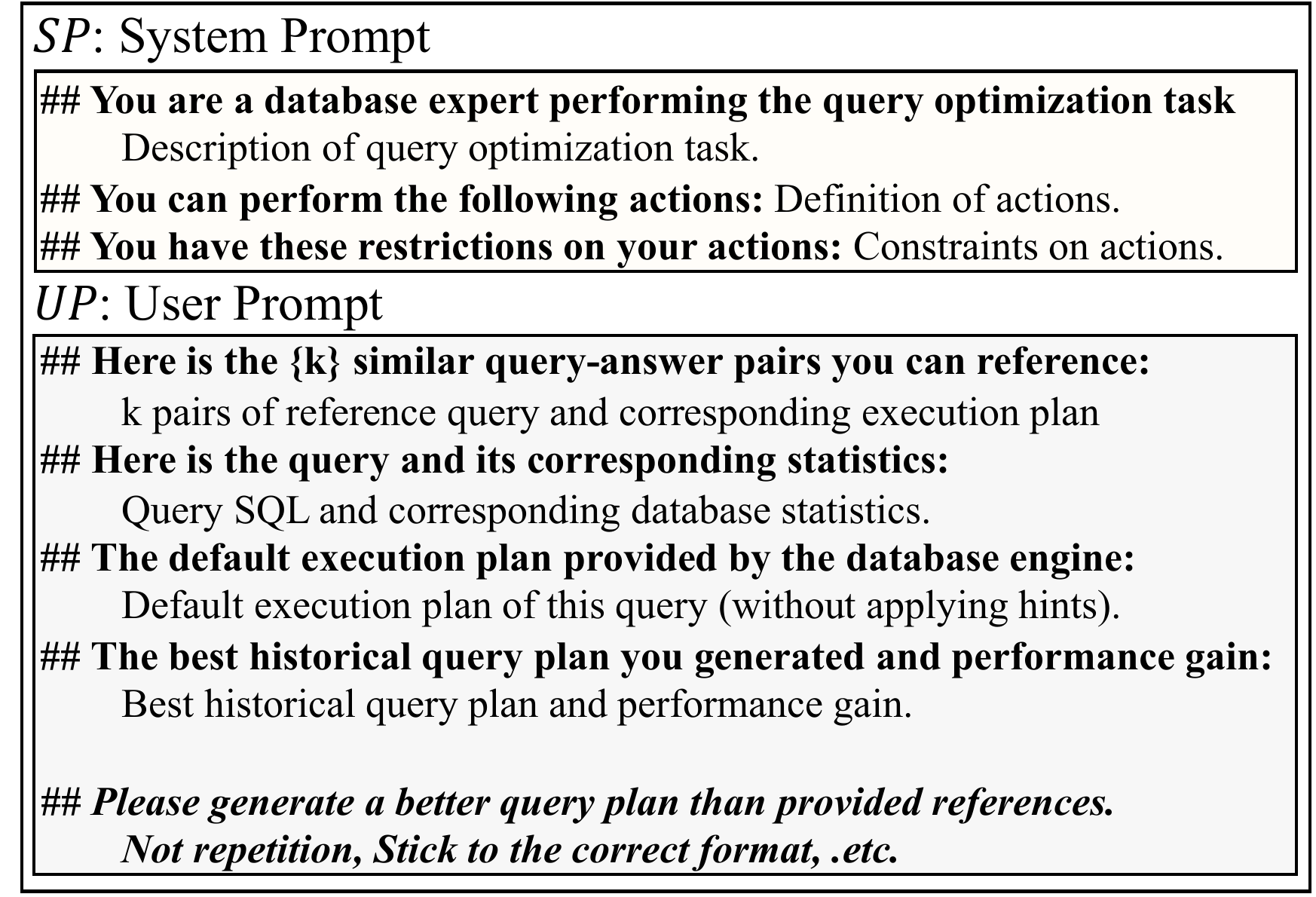}
    \caption{Dynamic prompt in {\oursystem}.}
    \label{fig:prompt_example}
\end{figure}

Figure~\ref{fig:prompt_example} illustrates the complete structure of the dynamic prompt \(\mathbf{x} = (\mathrm{SP}\,\|\,\mathrm{UP})\). It begins with a static system prompt, SP, which will not change. $\mathrm{UP}$ specifies the LLM’s role (e.g., database expert), the actions the LLM should perform, and the regulations of these actions in the context of our query optimization task. For a query~\(Q\), {\oursystem} generates a dynamic user prompt, UP, which initially includes the \(k\) most similar queries and their execution plans as references. These records provide the LLM with strongly correlated contextual information, as they represent the queries most similar to \(Q\) in the embedding space derived from the SQL text. The LLM can learn from these existing plans, even those for other queries, to better construct a query hint \(h\) for the input query \(Q\). Finally, the query statistics \(stats\), containing table cardinalities and filter selectivities, are appended to the prompt as additional context.

The following part of the user prompt contains the currently generated best query plan for query \(Q\) and the corresponding performance gain \(\eta\) relative to the baseline execution record (i.e., \(Q\) executed without any hint \(h\) during the \textit{Initialization} stage). These two items serve as heuristics for the LLM to understand the status (i.e., relative performance gain) of the current hint \(h\) and to target an optimized hint \(h^\ast\). At the end of the user prompt, we add expectations and regulations for the LLM, including (1) generating a better hint, (2) avoiding copying the plan provided by the execution engine or its own historical hints, and (3) ensuring output in the correct format so that the hint constitutes a valid plan.

\subsection{Self-evolving Feedback Paradigm}
\label{sec:Self-evolving Feedback Paradigm}

The middle part of Figure~\ref{fig:rag-framework} depicts the self‑evolving feedback paradigm of {\oursystem}. This paradigm enables the LLM to learn from historical execution records of the same query \(Q\), thereby enhancing the current iteration’s generated hint \(h\). For example, when we run the JOB 20a query~\cite{leis2015good} with {\oursystem}, the partial default query plan obtained from PostgreSQL during the \textit{Preparation} stage is \texttt{NestedLoop(k mk)} \(\rightarrow\) \texttt{NestedLoop(k mk cc)}. Then, in the \textit{Generation} stage (first iteration), {\oursystem} generates a hint for the plan \texttt{NestedLoop(mk k)} \(\rightarrow\) \texttt{NestedLoop(mk k t)}. In later iterations during the feedback process, the hint evolves to \texttt{HashJoin(mk k)} \(\rightarrow\) \texttt{NestedLoop(mk k cc)}. In this example, we observe that SEFRQO identifies two key insights: (1)~joining \texttt{mk} and \texttt{k} first is beneficial, and performing this join with a HashJoin while adjusting the join order yields better efficiency than a NestedLoop; and (2)~the nested loop join NestedLoop(mk k t) is efficient and should be retained. 

Algorithm~\ref{algorithm:Self-evolving Feedback Paradigm} illustrates the cross-iteration prompt optimization process (the straight line below line 8 indicates the separation between previous and current iterations). Here, we use $Prompt$ and $Prompt^\ast$ in place of the original notation $\mathbf{x}$ to denote the prompt from the previous iteration and the optimized prompt used in the current iteration, respectively. In the previous iteration part, we obtain the generated hint \(h\) by sending \(Prompt\) to LLM. Then, we execute query \(Q\) with \(h\) on the database engine to retrieve the resulting query plan and its execution time (lines 3–4). We then retrieve the historical best plan \({plan}^\ast\) and its execution time \(t^\ast\) from VecDB (line 5). Finally, we compare the current iteration’s result with the historical record, updating \({plan}^\ast\) and \(t^\ast\) if the current execution is superior (lines 6–8).

After that, we retrieve the baseline execution time \(t_o\) for query \(Q\) from VecDB (line 9). This time is stored during the \textit{Initialization} stage, representing the query executed without applying any hint. We then compute the performance gain \(\eta\) of the historical best execution time \(t^\ast\) over the baseline, \(\eta = (t_o - t^\ast)/t_o\) (line 10). Note that if \(t^\ast > t_o\), then \(\eta < 0\), indicating that the current best plan performs worse than the baseline. We merge the best plan \(\mathit{plan}^\ast\) and \(\eta\) into the user prompt \(UP\) (as in Section~\ref{sec:Prompts Design}) to provide useful information guiding LLM generation (line 11). Finally, we construct the optimized prompt \(Prompt^\ast\) by concatenating the static system prompt \(SP\) and the dynamically updated user prompt \(UP\) (line 12).

This optimized prompt, \(Prompt^\ast\), prevents the LLM from repeating the performance pitfall encountered with the previously generated hint \(h\) and continuously inspires the LLM to produce an optimal hint \(h^\ast\). As the pipeline runs iteratively, feedback from hints generated in earlier iterations drives the self‑evolving process. Meanwhile, the reference queries stored in VecDB are continuously updated, further enhancing the effectiveness of the references. These components iteratively optimize the prompt \(Prompt^\ast\), motivating the LLM to generate an optimized hint \(h^\ast\) that minimizes the execution latency for query \(Q\).


\begin{algorithm}[t]
\caption{Prompt Optimization in our RAG System}
\label{algorithm:Self-evolving Feedback Paradigm}
\begin{algorithmic}[1]
\State \textbf{Input:} Input query $Q$, Previous prompt $Prompt$ 
\State \textbf{Output:} Optimized prompt for current iteration $Prompt^\ast$
\State $h \gets \text{LLM}(Prompt)$ \Comment{Previous generated hint}
\State $plan, t \gets \text{ExecEngine}(Q, h)$
\Comment{Previous execution record}
\State  $plan^\ast, t^\ast \gets GetHistoricalBestExecutionTime(Q)$
\If{$t < t^\ast$}
    \State $plan^\ast \gets plan$ \Comment{Update historical best  plan}
    \State $t^\ast \gets t$ \Comment{Update historical best  execution time}  
\EndIf
\Statex \vspace{-10pt}\hspace{\algorithmicindent}\rule{0.93\linewidth}{0.4pt}
\State $t_o \gets \text{GetBaselineExecutionTime}(Q)$ 
\Comment{Retrieve execution time from record stored in the initialization stage from VecDB}
\State Compute the performance gain:
\(
\eta \gets \frac{t_o-t^\ast}{t_o}
\)
\State \(\mathrm{UP} \gets \mathrm{UP} \cup \{(plan^*,\,\eta)\}\) \Comment{Insert historical best plan and its performance gain into user prompt}
\State $Prompt^\ast \gets \text{ConstrcutPrompt}(SP, UP)$ 
\State \textbf{Return} $Prompt^\ast$
\end{algorithmic}
\end{algorithm}

\section{Evaluation}

We first introduce our experimental setup in Section~\ref{sec:Experimental Setup}. Then, we evaluate {\oursystem} using different benchmarks and multiple {\llm}s to address the following questions: (1) What is the performance gain over PostgreSQL and current LQOs in static and dynamic scenarios (Section~\ref{sec:MainResult})? (2) How effective is {\oursystem}'s self-evolving RAG feedback paradigm (Section~\ref{sec:self-evolving feed-
back Paradigm})? (3) What is the effect of SFT and {\qgrpo} on the quality of generated plans (Sections~\ref{sec:Fine-tuning Effect on the Starting Points})?  (4) What is the overhead of {\oursystem} and which factors influence it (Section~\ref{sec:overhead})? (5) How do different system settings affect the performance of {\oursystem} (Section~\ref{sec:Ablation Studies on Different Settings})? (6) How does {\oursystem} perform under various scenarios (Section~\ref{sec:extra-investigation})? {\oursystem} is open-source~\footnote{https://github.com/ihanwen99/SEFRQO}.

\subsection{Experiment Setup}
\label{sec:Experimental Setup}
\noindent \textbf{Environment.}
Unless mentioned otherwise, we run all our experiments on one server with an AMD Ryzen 9700X CPU, 128 GB of RAM, and an NVIDIA A6000 GPU with 48GB of memory. We conducted the majority of our experiments on PostgreSQL (version 16.4) with the corresponding {\pghint} version, and employed MySQL (version 8.4)~\cite{mysql} to assess {\oursystem}’s cross-DBMS capability. We use the Milvus~\cite{wang2021milvus} vector database for our RAG process. We evaluated {\oursystem} using $6$ local {\llm}s: LLaMA3.1-3B (\textbf{L3B}), LLaMA3.1-8B (\textbf{L8B}), their two SFT versions (\textbf{LS3B}, \textbf{LS8B}), and their two SFT-plus-{\qgrpo} versions (\textbf{LSG3B}, \textbf{LSG8B}).
We warm up the machine before running the experiments to mitigate caching issues. We use \textit{SentenceTransformer}\footnote{Paraphrase-MiniLM-L6-v2 Model.} as the embedding model to embed the plain SQL to a vector representation in our retrieval process.


\noindent \textbf{{\oursystem} Variants.} We evaluate three variants of {\oursystem}: (1) the default configuration, which leverages the full-plan hint in combination with reference queries retrieved from the vector database; (2) a variant that utilizes only the join order component of the hint (the “Leading” clause as detailed in Section~\ref{subsec:hint}), referred to as \textbf{JO}; and (3) a variant that relies exclusively on the historical execution records of the current query \( Q \) itself, without incorporating any reference queries from the RAG, referred to as \textbf{NR}. For clarity, we denote models using these configurations with suffixes, e.g., an LSG3B model employing the NR variant is denoted as \textbf{LSG3B-NR}. Unless mentioned, the number of retrieved reference queries is fixed at \( k = 1 \). The effect of varying \( k \) is investigated in Section~\ref{sec:Ablation Studies on Different Settings}.

\noindent \textbf{LQO Baselines.} We compare {\oursystem} with two state-of-the-art LQOs: Bao~\cite{marcus2021bao} and Balsa~\cite{yang2022balsa}. Bao is a steering LQO that tunes query plans generated by traditional optimizers through a learned model that selects different hint sets to enable/disable some operators like Hash Join, Index Scan, etc. Balsa is a generative LQO that learns a model to construct the query plan from scratch. 
For LLM-based LQOs, due to the lack of available complete code for LLM-QO~\cite{llm4qo-1} and LLMOpt~\cite{llm4qo-2}, we were unable to successfully reproduce their experiments. The only available LLM-based LQO implementation is LLMSteer~\cite{akioyamen2024unreasonable}, which we use as our primary baseline for evaluation in this category.  LLMSteer uses a text-embedding LLM to generate an embedding vector for the input SQL query and performs predictions using simple neural networks.
To have a fair comparison, we follow the best configurations of all baselines as mentioned in their papers\footnote{Balsa disables Bitmap Scan, Tid Scan, and Genetic Query Optimizer.}. 


\begin{table*}[tbp]
  \centering
  \caption{Performance gain in static scenarios \(\uparrow\).}
  \label{tab:performancegain_static}
  \begin{tabular}{cc|*{8}{c}}
    \hline
    Workload & Performance Gain & Balsa   & Bao     & L3B    & L8B    & LS3B   & LS8B   & LSG3B & LSG8B  \\ 
    \hline
    \multirow{2}{*}{CEB} 
                & Overall  & \textbf{37.57\%} & 22.60\% & 7.46\% & 21.78\% & 53.93\% & \textbf{60.04\%} & 50.13\% & 48.36\% \\
                & Filtered & \textbf{59.22\%} & 30.13\% & 17.93\% & 47.56\% & 56.76\% & \textbf{65.05\%} & 55.34\% & 61.28\% \\ 
    \hline
    \multirow{2}{*}{Stack} 
                & Overall  & 67.17\% & \textbf{76.90\%} & 5.54\%  & 15.21\% & \textbf{40.10\%} & 26.77\% & 39.62\% & 32.58\% \\
                & Filtered & \textbf{89.83\%} & 82.85\% & 9.16\%  & 37.38\% & 43.29\% & 87.68\% & 42.77\% & \textbf{93.57\%} \\ 
    \hline
  \end{tabular}
\end{table*}

\noindent \textbf{Evaluation Metrics.} We focus mainly on the following metrics: (1)~\underline{\textit{Performance Gain in Query Execution Latency:}} In this metric, we measure the performance gain over PostgreSQL. We compute this performance gain in two modes. The first mode is \textit{Overall Performance Gain}, where we calculate the weighted sum of the performance gains in ``all'' the workload queries using {\oursystem} (or any LQO) over using PostgreSQL. The weight of each query is the ratio between its corresponding execution time and the total time of all workload queries when running with PostgreSQL. The second mode is \textit{Filtered Performance gain}, where the gain is calculated the same way, yet over the queries that have been accelerated ``only''. 
(2)~\underline{\textit{Relative Execution Time (RET):}} In this metric, we compute the ratio between the overall execution time of all the queries using {\oursystem} (or any LQO) and using PostgreSQL.
(3)~\underline{\textit{Inference and RAG Retrieval Latency:}} These metrics measure the latency overhead of the LLM inference and the RAG retrieval from the vector database. (4)~\underline{\textit{Homogeneous Rate (HR):}} We assess the diversity of LLM-generated hints by calculating the ratio between (i) the number of invalid hints or hints identical to those plans generated by the PostgreSQL and (ii) the total number of generated hints. A higher \textit{HR} indicates that the model is either generating invalid hints (rejected by the database as plans) or producing repetitive hints that copy the execution engine’s own plans. Note that we do not distinguish between invalid hints and the execution engine's default plans because invalid hints result in the same default plans, making them indistinguishable in practice. Evaluating the diversity of hints generated by LLMs is essential because only distinct hints can effectively influence the database’s execution behavior. 

For all evaluation metrics, we use $\uparrow$ to denote “the higher, the better,” and $\downarrow$ to denote “the lower, the better.” These symbols will be consistently used across all figures and tables.

\noindent \textbf{Benchmarks.} We use the \textit{Join Order Benchmark (JOB)}~\cite{leis2015good} workload to (1)~fine-tune the LS3B, LS8B, LSG3B, and LSG8B models and (2)~pre-populate the retrieval corpus in the RAG system with query-answer examples during the preparation phase. Subsequently, we evaluate {\oursystem} on two workloads: the \textit{Cardinality Estimation Benchmark (CEB)}~\cite{flowloss} and \textit{Stack}~\cite{marcus2021bao}. Both JOB and CEB are constructed on the IMDB movie database schema and data, with CEB containing more complex query patterns and a wider range of cardinality estimation challenges than JOB. The Stack workload is derived from the Stack Overflow dataset; while its query style bears similarity to the IMDB-based workloads, it features a greater number of slow queries. 
Additionally, \textit{TPC-DS benchmark}~\cite{tpcds} is used to test {\oursystem} on  complex workloads. We use a scale factor of 4, consistent with prior work~\cite{chen2023loger}. For CEB, we use two queries per template; for Stack, five queries per template; and for TPC-DS, one query per template. To prevent poorly optimized plans from excessively blocking the benchmarking, we enforce the following timeout thresholds for different benchmark: 10s for the JOB and CEB workloads, 30s for Stack, and 60s for TPC-DS.



\subsection{Query Execution Latency Evaluation}\label{sec:MainResult}
In this section, we focus on the \textit{performance gain in query execution latency} for each query during a 50-iteration run of {\oursystem}. We evaluate on both static and dynamic scenarios. In static scenarios, we keep using the same workload for all iterations during the online stage. The first static scenario employs the CEB workload, which is used to test {\oursystem} under a different setup while retaining the same schema and data distribution. The second static scenario employs the Stack workload, which is entirely different from the JOB workload that is used for fine-tuning, and is used to verify {\oursystem}'s cross-domain transfer capability (i.e., workload switch). 
The dynamic scenario comprises a mixture of the CEB, JOB, and Stack workloads, alternating across different episodes. For each episode, one workload is selected randomly and executed for a number of iterations. We set 10 episodes and randomly allocate iterations to each episode, enforcing a total of 50 iterations.

\begin{figure}[htbp]
  \centering
  \subfloat[LS8B (CEB)]{
    \includegraphics[width=0.47\linewidth]{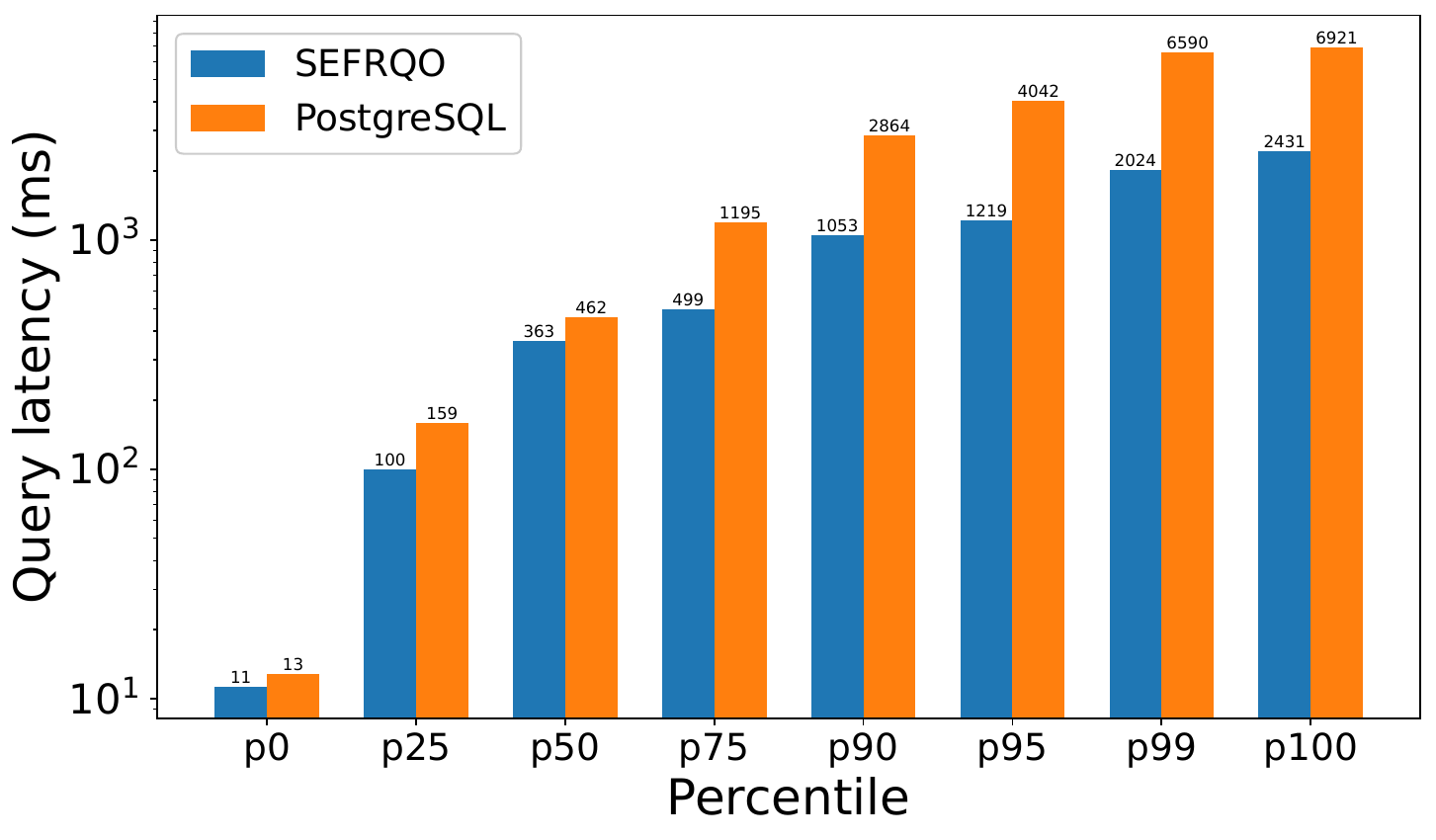}
    \label{fig:percentile-ceb-ls8b}
  } 
  \hfill
  \subfloat[LSG8B (CEB)]{
    \includegraphics[width=0.47\linewidth]{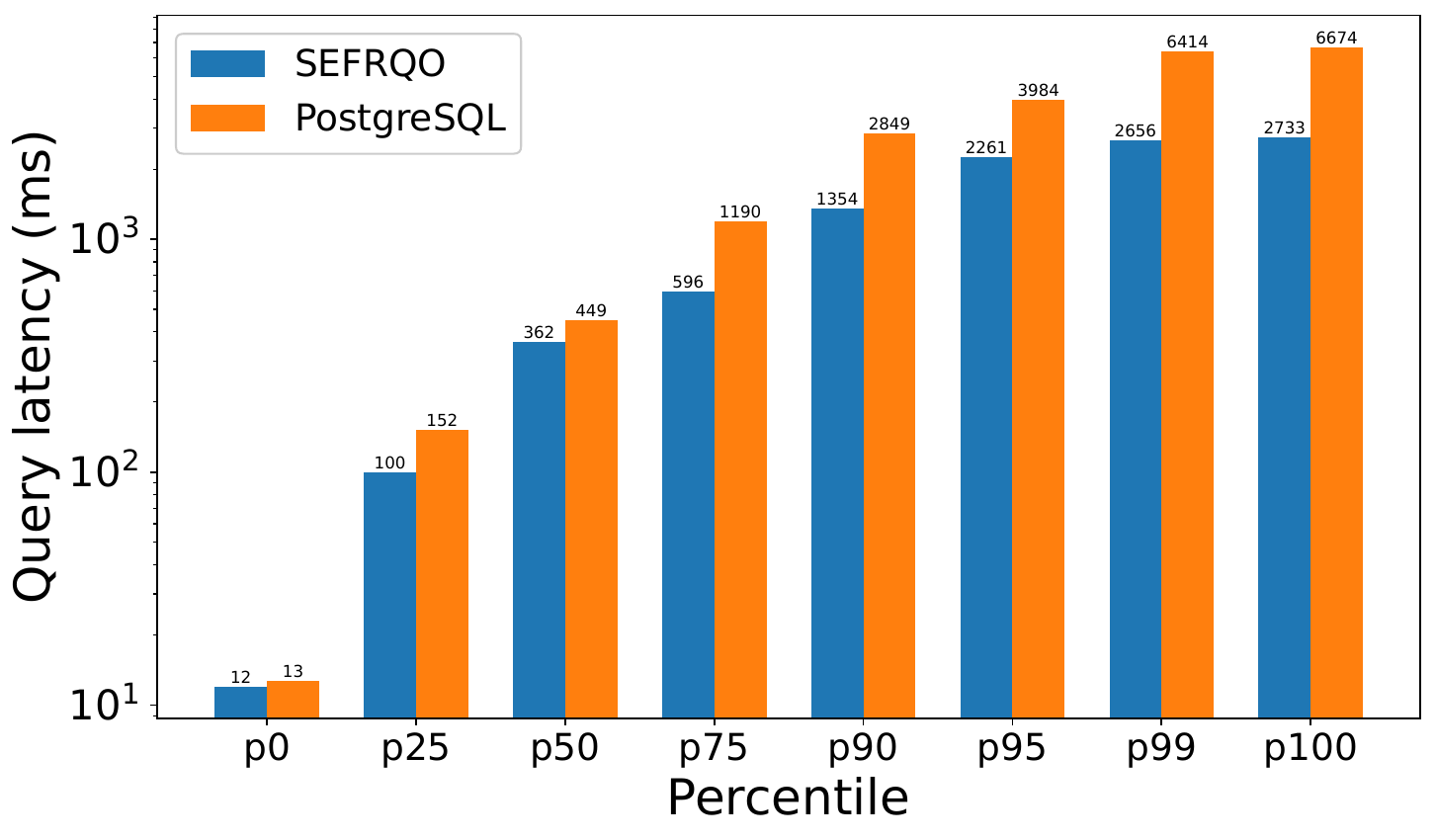}
    \label{fig:percentile-ceb-lsg8b}
  }
  \label{fig:percentile-ceb-all}
  \subfloat[LS8B (Stack)]{
    \includegraphics[width=0.47\linewidth]{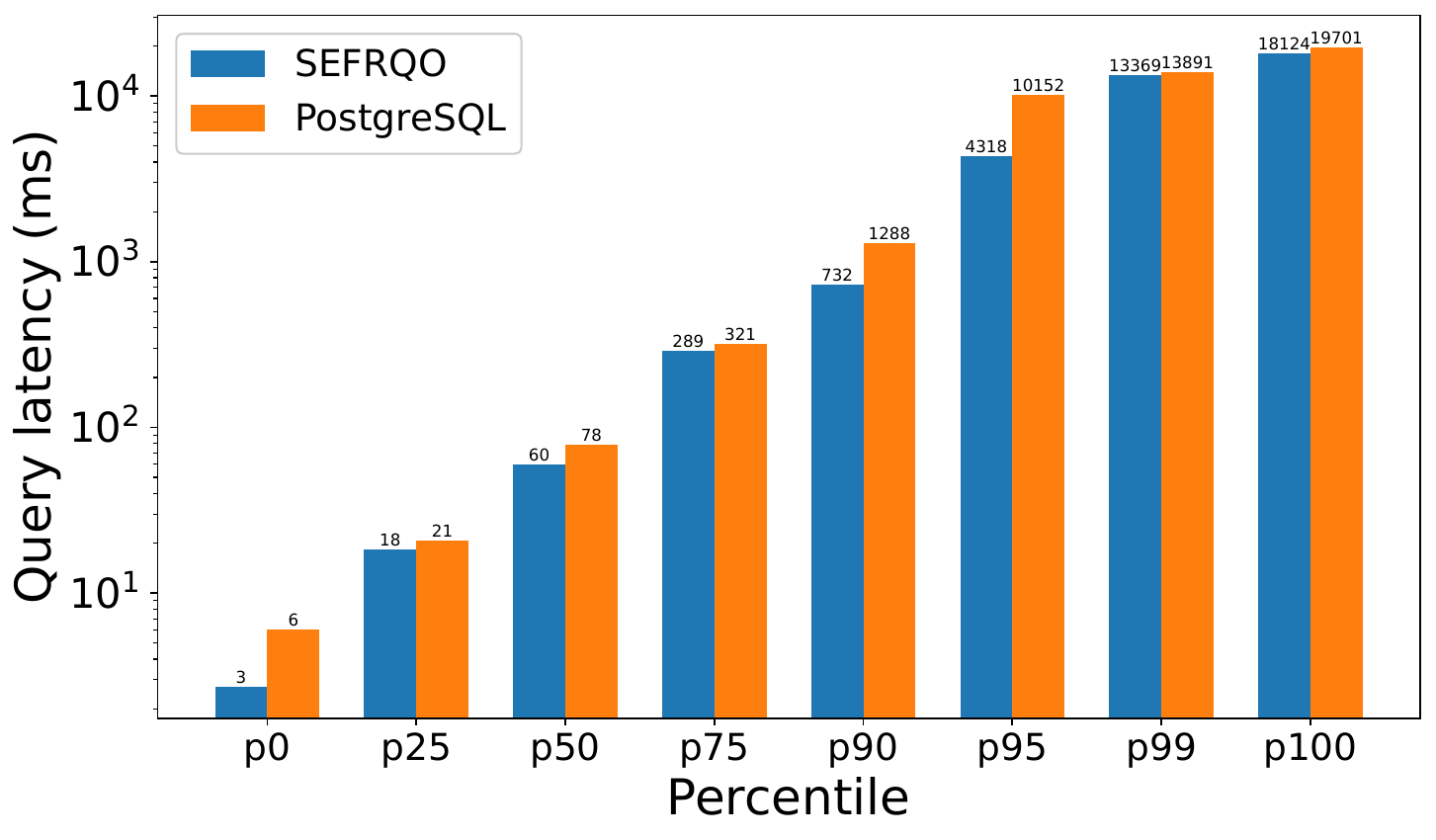}
    \label{fig:percentile-stack-ls8b}
  }
  \hfill
  \subfloat[LSG8B (Stack)]{
    \includegraphics[width=0.47\linewidth]{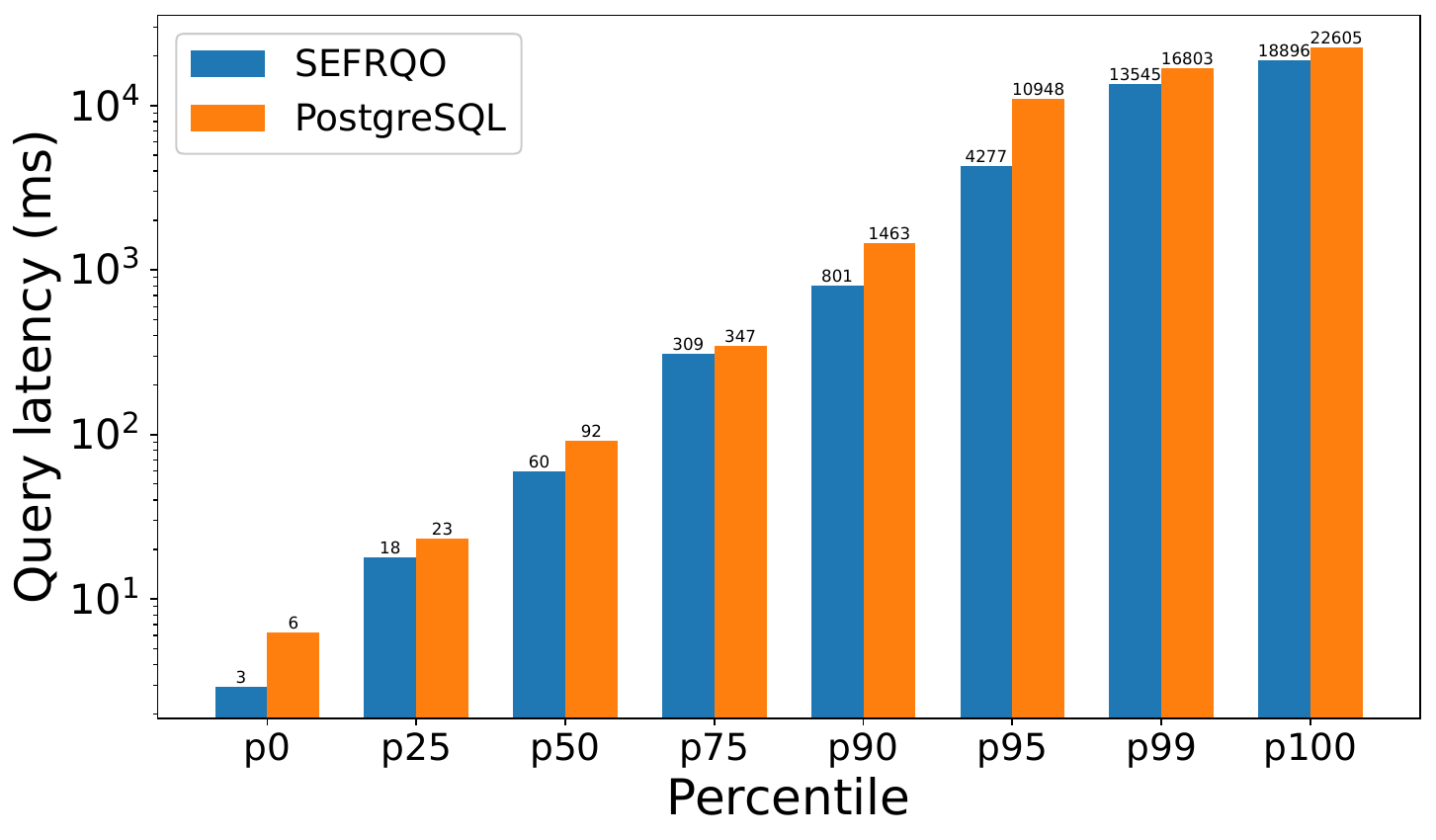}
    \label{fig:percentile-stack-lsg8b}
  }
  \caption{Query latency distributions on CEB and Stack.}
  \label{fig:percentile}
\end{figure}

\begin{figure*}[t]
  \centering
  \begin{minipage}[b]{0.45\textwidth}
    \centering
    \includegraphics[width=0.8\textwidth]{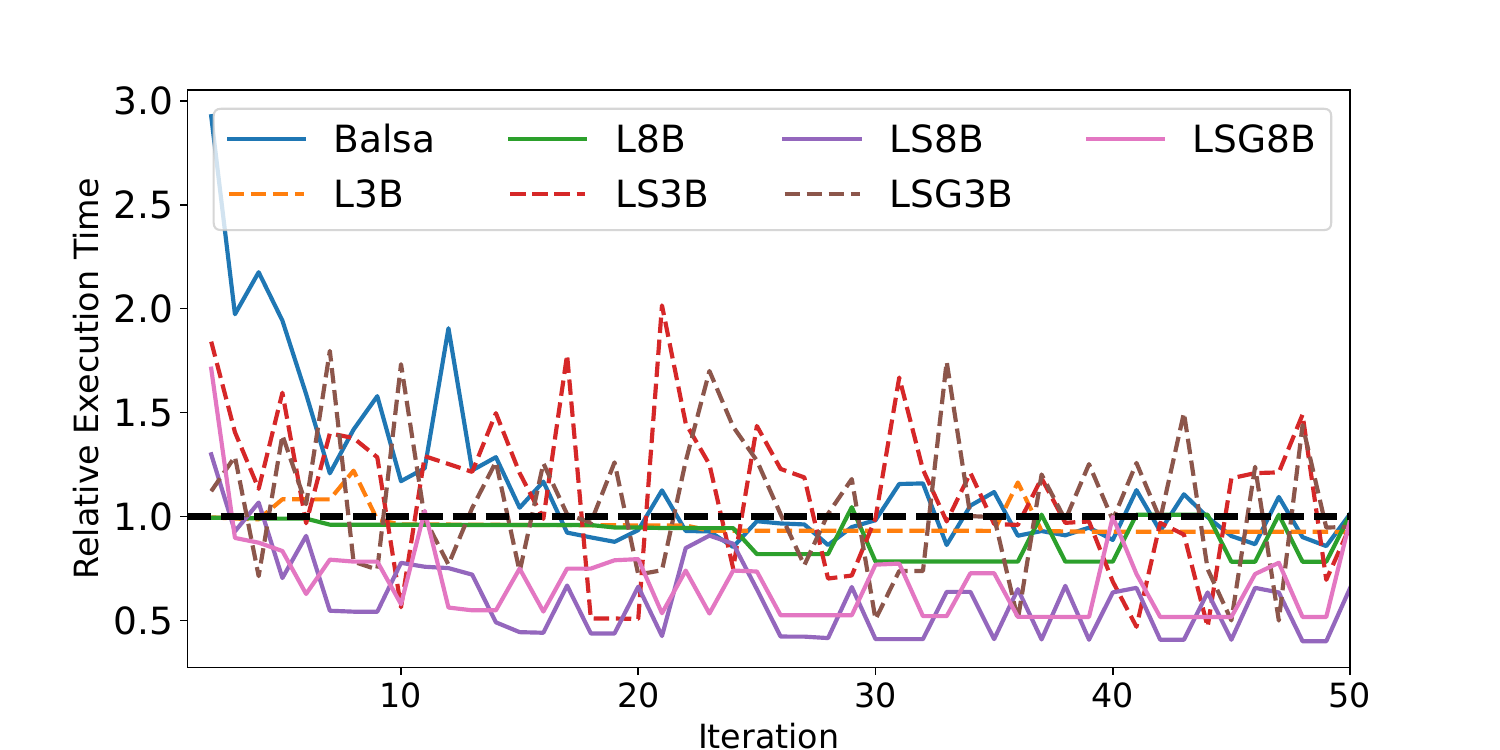}
    \caption{RET on CEB workload across iterations $\downarrow$.}
    \label{fig:ret_ceb}
  \end{minipage}
  \hfill
  \begin{minipage}[b]{0.45\textwidth}
    \centering
    \includegraphics[width=0.8\textwidth]{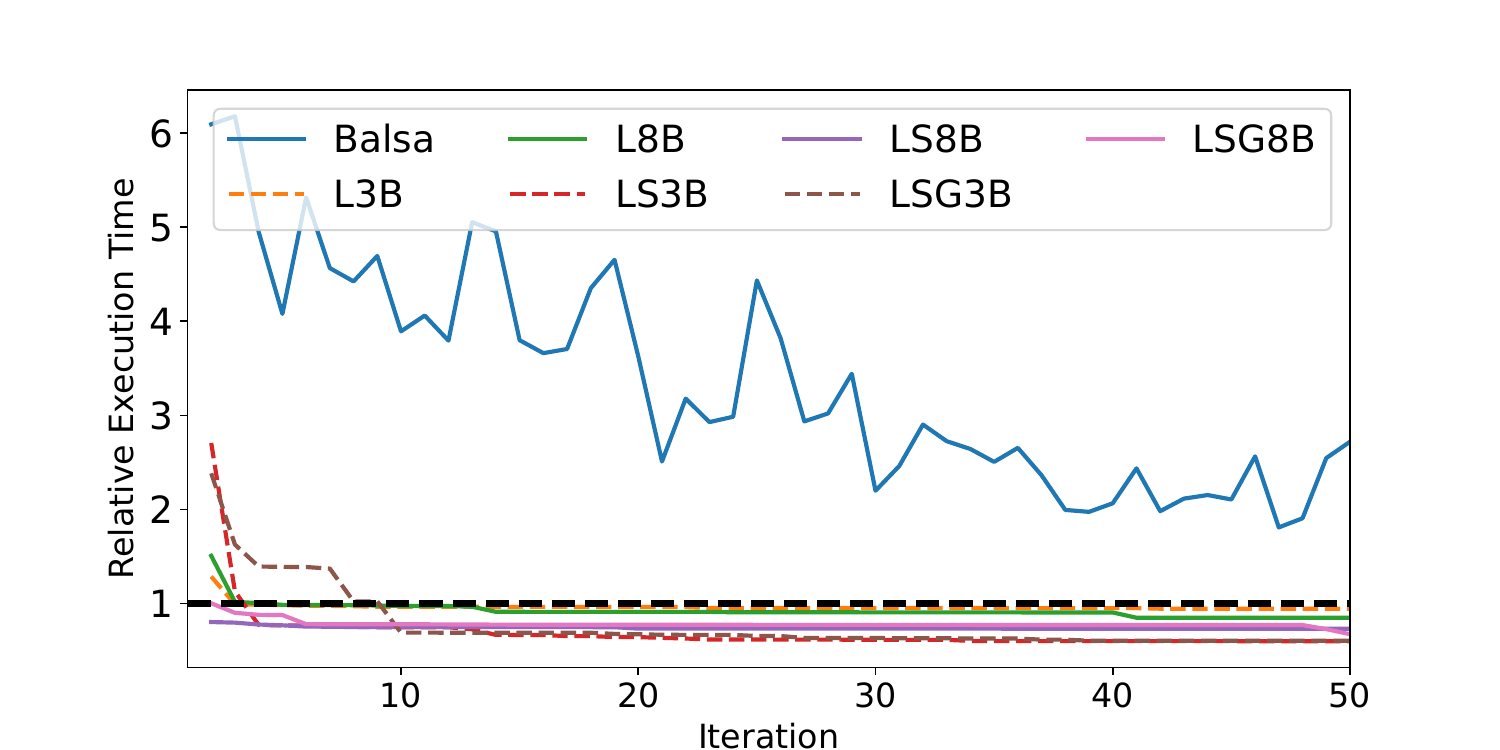}
    \caption{RET on Stack workload across iterations $\downarrow$.}
    \label{fig:ret_stack}
  \end{minipage}
\end{figure*}

\noindent \textbf{Performance on Static Workloads.} 
Table~\ref{tab:performancegain_static} presents the performance gains achieved on static workloads. The comparison baseline is PostgreSQL's default query plan. For LQOs, both Balsa and Bao are trained and evaluated on the same workload (either CEB or Stack). On the CEB workload, {\oursystem} with the LS8B model exhibits superior performance relative to the other models, and it outperforms LQOs. This result also suggests that fine-tuning is crucial, as the vanilla models (L3B and L8B) show significantly less improvement compared with models that have undergone supervised fine-tuning (LS3B and LS8B). Interestingly, a slight performance degradation is observed when applying {\qgrpo} after the SFT process. This is because SFT-based LLMs benefit the most (i.e., leverage the maximum of the fine-tuned knowledge) when they get fine-tuned and tested on similar workloads, such as JOB and CEB. However, when {\oursystem} confronts an unseen workload that differs significantly from the fine-tuning workload (as shown right below with the Stack workload) or when the RAG support is disabled (as shown in Section~\ref{sec:Fine-tuning Effect on the Starting Points}), SFT-plus-{\qgrpo} models demonstrate their advantages that stem from the exploratory nature of reinforcement fine-tuning, which enhance the reasoning capabilities of larger LLMs to handle such challenging situations. On Stack workload, both the SFT and SFT-plus-{\qgrpo} models exhibit good performance. Even though the LLMs have never seen this workload before, the filtered performance gains for LS8B and LSG8B are comparable to those of Balsa or outperform it. Notably, these LQOs are trained on Stack for 50 iterations. This scenario also demonstrates the remarkable knowledge-transfer capabilities of {\oursystem} with the SFT and SFT-plus-{\qgrpo} models, as evidenced by the substantial performance gains they achieve over their vanilla counterparts.

Figure~\ref{fig:percentile} compares the query latency distributions on both CEB and Stack workloads. {\oursystem} consistently outperforms PostgreSQL across all percentiles, indicating that it benefits the majority of queries and steadily reduces their latency. In particular, on the CEB workload, {\oursystem} is especially effective at optimizing long-tail (high-percentile) queries, achieving approximately 2.5× speedup.

\noindent\underline{Comparison against LLM-based LQOs.} As mentioned in Section~\ref{sec:Experimental Setup}, we experiment with LLMSteer~\cite{akioyamen2024unreasonable} as a representative for LLM-based LQOs. However, it is not fully integrated into a database engine and requires workloads to be preprocessed into a specific format (CSV files containing the SQL file name, raw SQL text, a mapping between hint lists and runtime lists, and a special representation of the query plan), without publicly available source code for that preprocessing. The authors only provided preprocessed files for the JOB and CEB workloads. Therefore, we only reproduced LLMSteer under the same setup as the first static scenario: training on the JOB workload and evaluating on the CEB workload. LLMSteer achieves an Overall Performance Gain (i.e., improvement across all queries) of $45.14\%$, while {\oursystem} achieves $60.04\%$. Although LLMSteer’s embedding-based generation and specialized architecture offer faster inference, they result in lower overall performance.

\noindent \textbf{Performance on Dynamic Scenario.} 
Table~\ref{tab:performancegain_dynamic} shows the capability of {\oursystem} to handle changing workloads. Notably, most existing LQOs do not support dynamic workloads, as their model architectures are typically tied to the schema of the training workload. However, since Bao can accommodate various schemas, we include it in this experiment. In general, larger LLMs yield better results, and SFT is crucial for enhancing performance. Since approximately two-thirds of the workloads originate from the IMDB schema (i.e., JOB and CEB workloads), the dominance of familiar patterns may diminish the effectiveness of SFT-plus-{\qgrpo} models, which tend to perform better on substantially different workloads such as Stack. Additionally, we can see that Bao underperforms compared to {\oursystem}, and exhibits a significant performance drop relative to its results in static settings. These results demonstrate the superiority of {\oursystem} in handling dynamic workloads, effectively eliminating reliance and constraints associated with specific workloads.

\begin{table}[htbp]
  \centering
  \caption{Performance gain in dynamic scenario $\uparrow$.}
  \label{tab:performancegain_dynamic}
  \begin{tabular}{lcc}
    \hline
    Model & \multicolumn{2}{c}{Performance Gain} \\ \cline{2-3}
          & Overall & Filtered \\ \hline
    L3B    & 2.64\%  & 20.11\%  \\
    L8B    & 15.92\% & 54.83\%  \\
    LS3B   & 50.25\% & 58.65\%  \\
    LS8B   & \textbf{53.25\%} & \textbf{69.98\%}  \\ 
    LSG3B  & 33.88\% & 46.38\%  \\
    LSG8B  & 38.83\% & 51.38\%  \\ \hline
   Bao  & 14.82\% & 31.36\%  \\ \hline
  \end{tabular}
\end{table}

\subsection{Self-evolving Feedback Paradigm}\label{sec:self-evolving feed-
back Paradigm}

In this section, we shift our focus to the overall execution latency for all queries in each iteration. We use the \textit{RET} metric to measure the execution time ratio of {\oursystem} compared to PostgreSQL. 


Figure~\ref{fig:ret_ceb} shows the self-evolving process under the CEB workload. Among all evaluated models, {\oursystem} with LS8B exhibits the best performance, achieving the lowest \textit{RET} of 0.41. It also shows a trend and potential for continuous decrease with additional iterations. In contrast, {\oursystem} with vanilla models (L3B, L8B) shows only marginal performance gains throughout the entire process. Notably, {\oursystem} with L8B exhibits minor fluctuations, {\oursystem} with L3B remains largely unchanged. This behavior likely results from L3B consistently generating the same query hint, thereby failing to effectively utilize information from the dynamic prompt we construct. However, this phenomenon is mitigated with larger LLMs, which actively absorb input and generate more varied hints. 



This experiment further proves that domain-specific fine-tuning plays a crucial role in {\oursystem}. Both LS3B and LS8B outperform their vanilla counterparts; however, LS3B demonstrates notable instability compared to LS8B, with performance fluctuation observed several times across all iterations. This phenomenon may indicate that smaller LLMs with supervised fine-tuning exhibit higher instability than larger models in our self-evolving paradigm.


Now, let us switch the perspective to compare {\oursystem} with LQOs. Balsa achieves only limited performance improvements over the baseline (denoted by the black dashed line) within each iteration. Although Balsa achieves promising results in the previous evaluations (i.e., by gathering the best queries among all iterations, as shown in Table~\ref{tab:performancegain_static}), these improved queries are scattered across different iterations. This indicates that within 50 iterations, Balsa does not consistently generate high-performance query plans in the same iteration, which means Balsa requires more iterations to reach peak performance. This is more obvious in Figure~\ref{fig:ret_stack} when testing on the Stack workload: Balsa fails to outperform the baseline within our iteration range despite exhibiting a trend towards performance improvement over iterations.


As Figure~\ref{fig:ret_stack} shows, LLMs exhibit more stable evolving progress (i.e., lower variance and fewer fluctuations across iterations) on the Stack workload compared to their behavior on the CEB workload. This may originate from the fact that the LLMs are not fine-tuned on the Stack workload, causing them to be more cautious and reluctant to substantially change their behaviors in generating hints. Consequently, it enables a smaller model to outperform a larger model. As evidence, LS3B and LSG3B are unable to surpass LS8B on the CEB workload, but they ultimately achieve better performance than LS8B on the Stack workload. In this scenario, LSG3B with \(Q_{\mathrm{GRPO}}\) fine-tuning achieves the best performance.


\begin{figure}[tb]
    \centering  \includegraphics[width=0.8\linewidth]{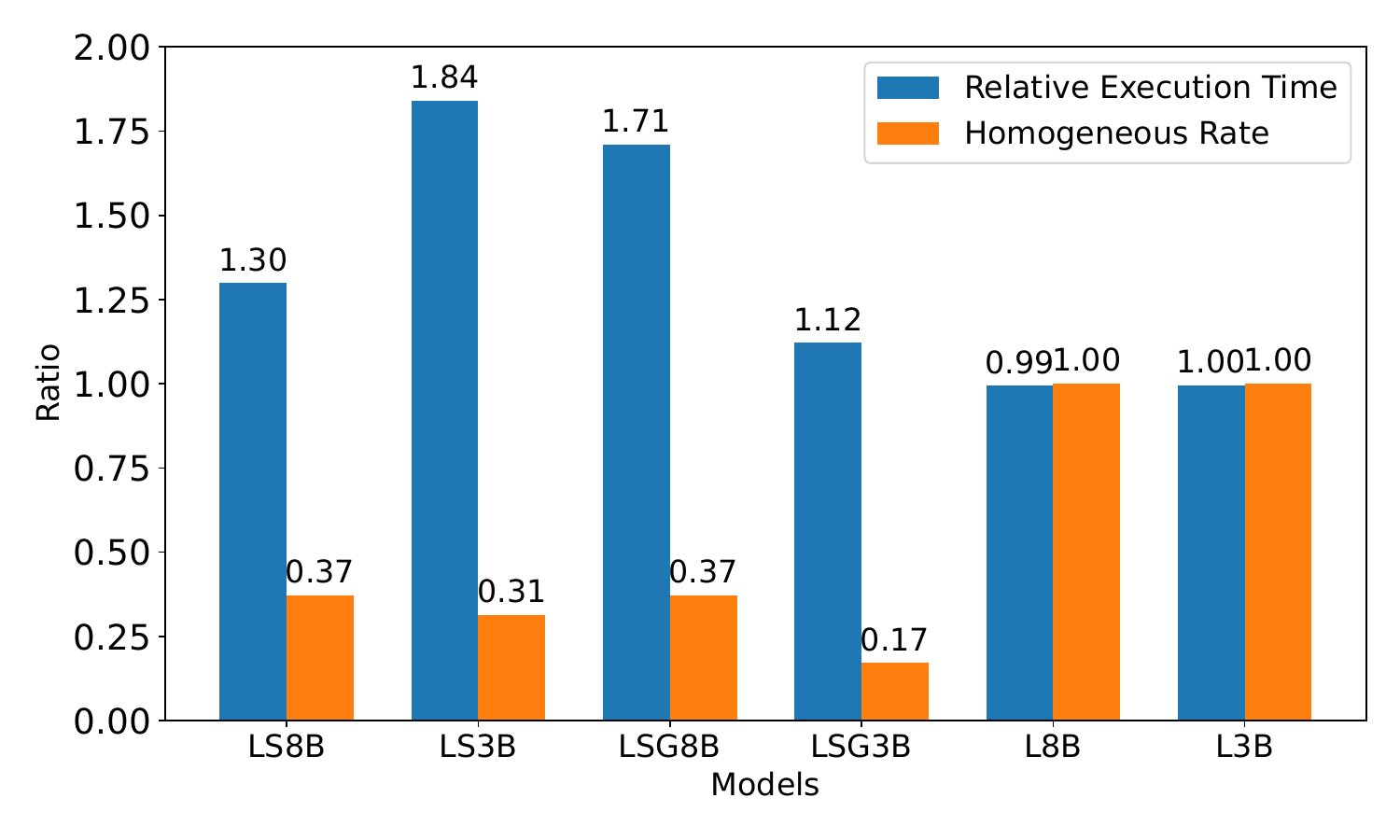}
    \caption{RET and HR for investigating fine-tuning effects (on the first iteration of {\oursystem}) $\downarrow$.}
    \label{fig:starting_points_data}
\end{figure}

\subsection{SFT and {\qgrpo} Effects}
\label{sec:Fine-tuning Effect on the Starting Points}
Previous experiments verify {\oursystem}'s collaborative capability by combining LLM fine-tuning with the self‑evolving paradigm. In this section, we focus on the plan generation ability based on LLM's internal reasoning, without involving the self-evolving paradigm introduced by RAG. We follow the initial setup, where dynamic prompts include only PostgreSQL execution records, excluding the best plan or performance gain, and evaluate using \textit{RET} and \textit{HR}.

Figure~\ref{fig:starting_points_data} shows the results for different LLMs on the CEB static workload (as in Section~\ref{sec:MainResult}). Both vanilla LLMs (i.e., L8B and L3B) primarily replicate PostgreSQL’s default execution plans or repeatedly generate invalid hints and fall back to the PostgreSQL plan, yielding performance similar to PostgreSQL's plan (i.e., RET close to 1) and a high homogeneity rate (i.e., HR close to 1). In contrast, the fine-tuned LLMs produce more diverse plans (i.e., lower HR), favoring exploration over simply copying the input query plan. This indicates an improvement in the reasoning abilities. Even when the initial plan is suboptimal (i.e., higher RET), our self-evolving paradigm reduces the final RET (see Figures~7 and~8).


We observe the pronounced {\qgrpo} effect in the 3B models: from LS3B to LSG3B, both the \textit{HR} and \textit{RET} decrease further. This shows that {\qgrpo} enhances the models’ reasoning abilities, enabling 3B‑level models to generate more diverse and superior query hints. Notably, LSG3B even outperforms the larger LS8B model on both metrics. This demonstrates that a successful {\qgrpo} fine-tuning process can steer the model’s output preferences to achieve better task-specific performance compared to larger LLMs without {\qgrpo}. Given that {\qgrpo} is based on reinforcement learning, it introduces inherent uncertainty into LLM fine-tuning processes, which may require multiple fine-tuning runs to obtain a reliable model. For example, we show that the LSG8B model does not show improvement over LS8B. This may be attributed to the more complex reasoning nature of 8B models, which prevents them from demonstrating superiority without the self‑evolving RAG paradigm.

\subsection{Inference and RAG Retrieval Time}\label{sec:overhead}
This experiment evaluates the overhead incurred by LLM inference and RAG retrieval. We analyze these overheads for {\oursystem} using the LS3B model. The LLM inference time is approximately $236\,\text{ms}$ on an NVIDIA A6000 and $160\,\text{ms}$ on an NVIDIA A100. The average RAG retrieval latency is $0.7\,\text{ms}$. We find that the dominant overhead arises from LLM inference, which can be mitigated by using powerful GPUs. Furthermore, considering execution time savings (with {\oursystem}'s query plans for the CEB workload averaging $520.67\,\text{ms}$ compared to PostgreSQL's $922.86\,\text{ms}$), {\oursystem}'s end-to-end latency still remains lower for OLAP.

\begin{figure}[H]
    \centering  \includegraphics[width=\linewidth]{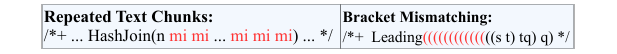}
    \caption{Two classical errors during LLMs' generations, unneeded output characters increase inference time.}
    \label{fig:wrong_egs}
\end{figure}

Inference time of LLMs is also positively correlated with output length. In our experiments, we observed two types of erroneous hint generation. Although PostgreSQL will reject such hints, we still incur a much longer hint generation. Figure~\ref{fig:wrong_egs} presents the two most classic errors. The first is \textbf{Repeated Text Chunks}, where the generated hint repeatedly includes the same table names multiple times. The second is \textbf{Bracket Mismatching}, in which excessive and unbalanced parentheses are generated. One direct implication is to incorporate a validation procedure during inference: when an unusable hint is generated, the process can be aborted promptly to reduce inference time.

\begin{figure}[t]
  \centering

  \begin{subfigure}[b]{0.48\columnwidth}
    \centering
    \includegraphics[width=1.1\linewidth]{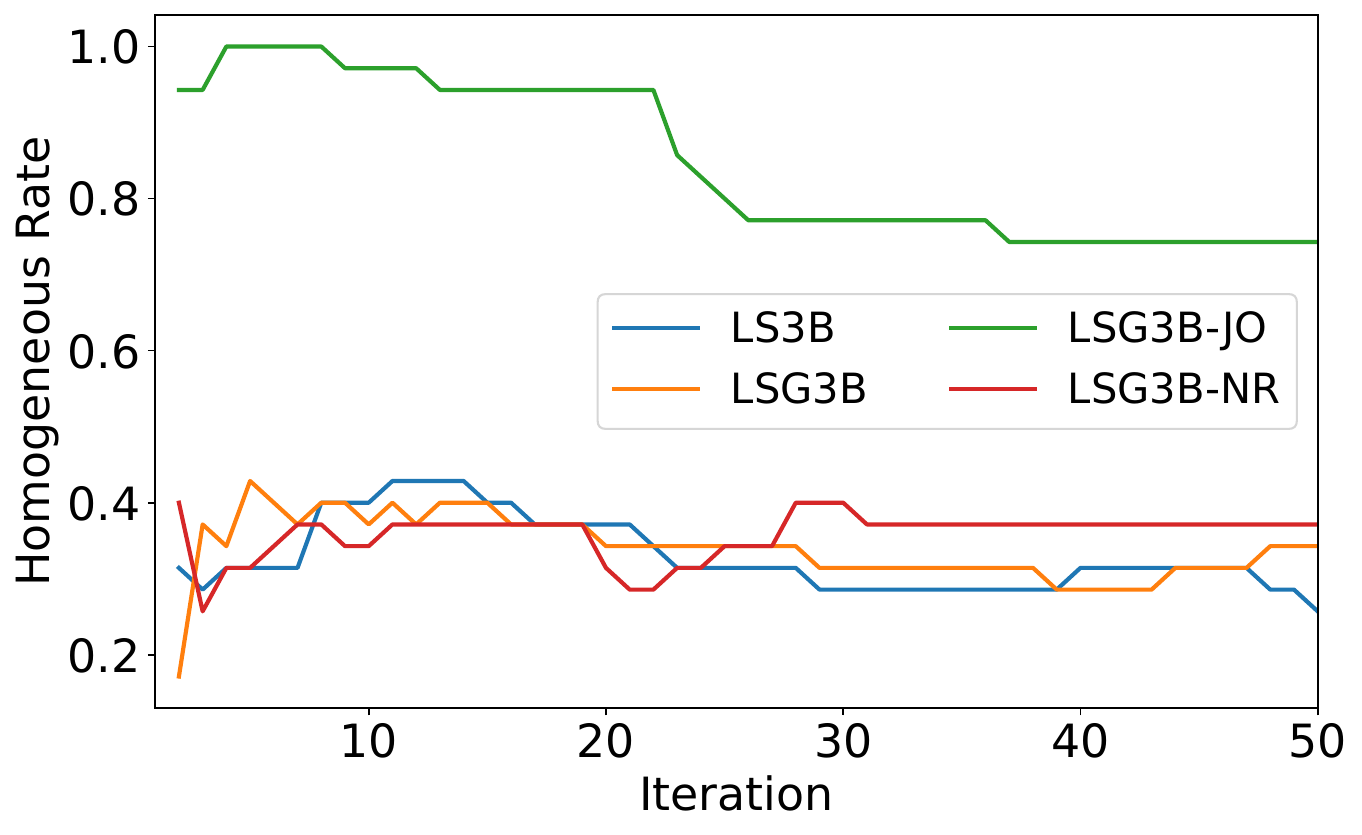}
    \caption{HR $\downarrow$.}
    \label{fig:hr_ceb_ablation}
  \end{subfigure}
  \hfill
  \begin{subfigure}[b]{0.48\columnwidth}
    \centering
    \includegraphics[width=1.1\linewidth]{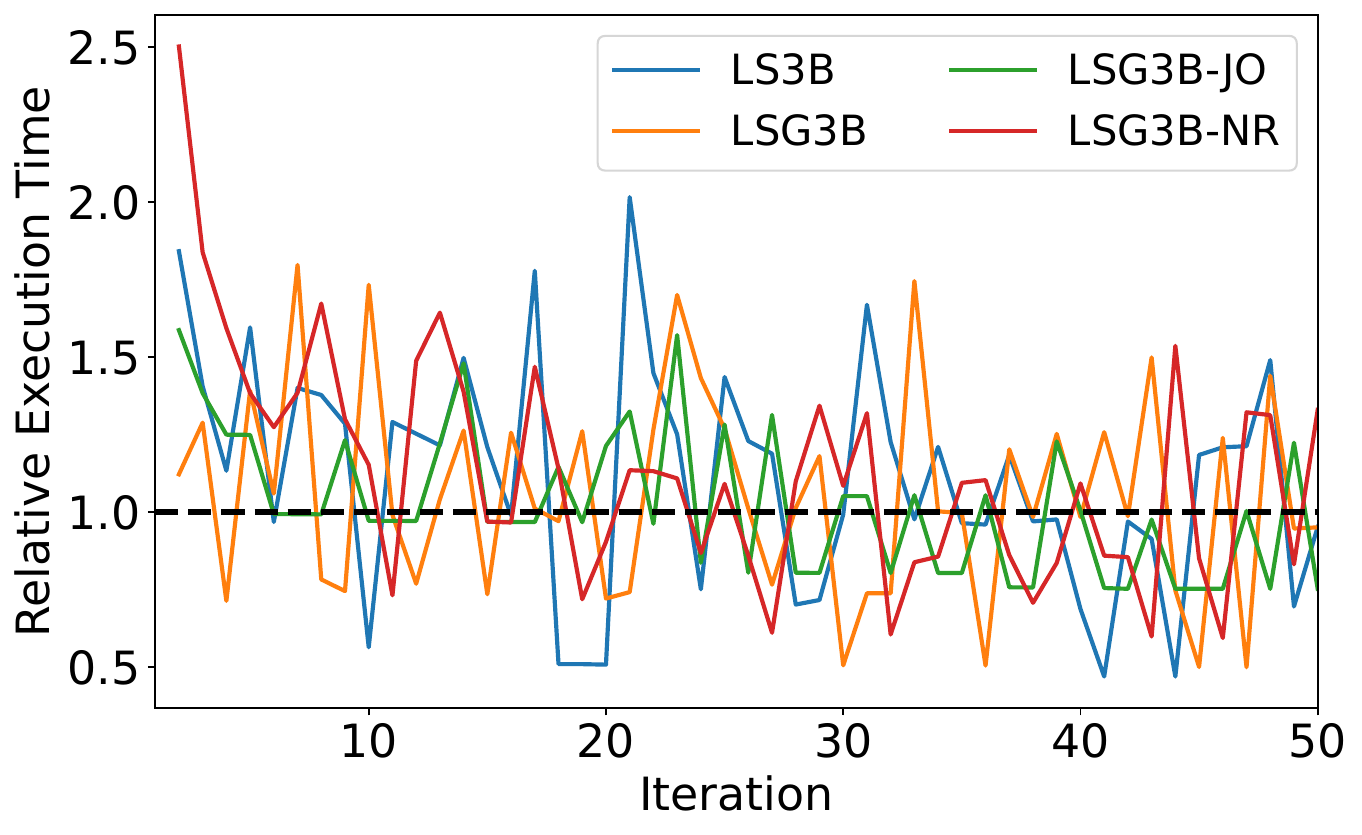}
    \caption{RET $\downarrow$.}
    \label{fig:ret_ceb_ablation}
  \end{subfigure}

  \caption{Evaluation on the static CEB workload.}
\end{figure}


  
\begin{table}[htbp]
  \centering
  \caption{Performance gain on static CEB workload $\uparrow$.}
  \label{tab:ablation_studies}
  \begin{tabular}{lrrrr}
    \toprule
    Performance Gain & LS3B & LSG3B & LSG3B-JO & LSG3B-NR \\
    \midrule
    Overall  & \textbf{53.93\%} & \textbf{50.13\%} & 24.81\% & 41.26\% \\
    Filtered & \textbf{56.76\%} & \textbf{55.34\%} & 48.65\% & 52.61\% \\
    \bottomrule
  \end{tabular}
\end{table}

\subsection{Ablation Studies}
\label{sec:Ablation Studies on Different Settings}
In this section, we focus on 3B models and evaluate several configurations to assess their impact on {\oursystem}.

\noindent \textbf{Generation Mode.}
Figure~\ref{fig:hr_ceb_ablation} depicts the \textit{HR} for each iteration. We observe that LSG3B-JO achieves a higher \textit{HR} (approximately 70\%) than the other configurations. These results indicate that when LLM in {\oursystem} is constrained to generate only a join-order hint, it mainly reproduces or approximates the default plan provided by the execution engine and is reluctant to propose new hints to improve performance. Table~\ref{tab:ablation_studies} presents the performance gains over the baseline, comparing LSG3B with LSG3B-JO. The results demonstrate that the full-plan hint generation mode is substantially more effective than the \textbf{JO} mode.


\begin{figure}[tb]
    \centering  \includegraphics[width=0.8\linewidth]{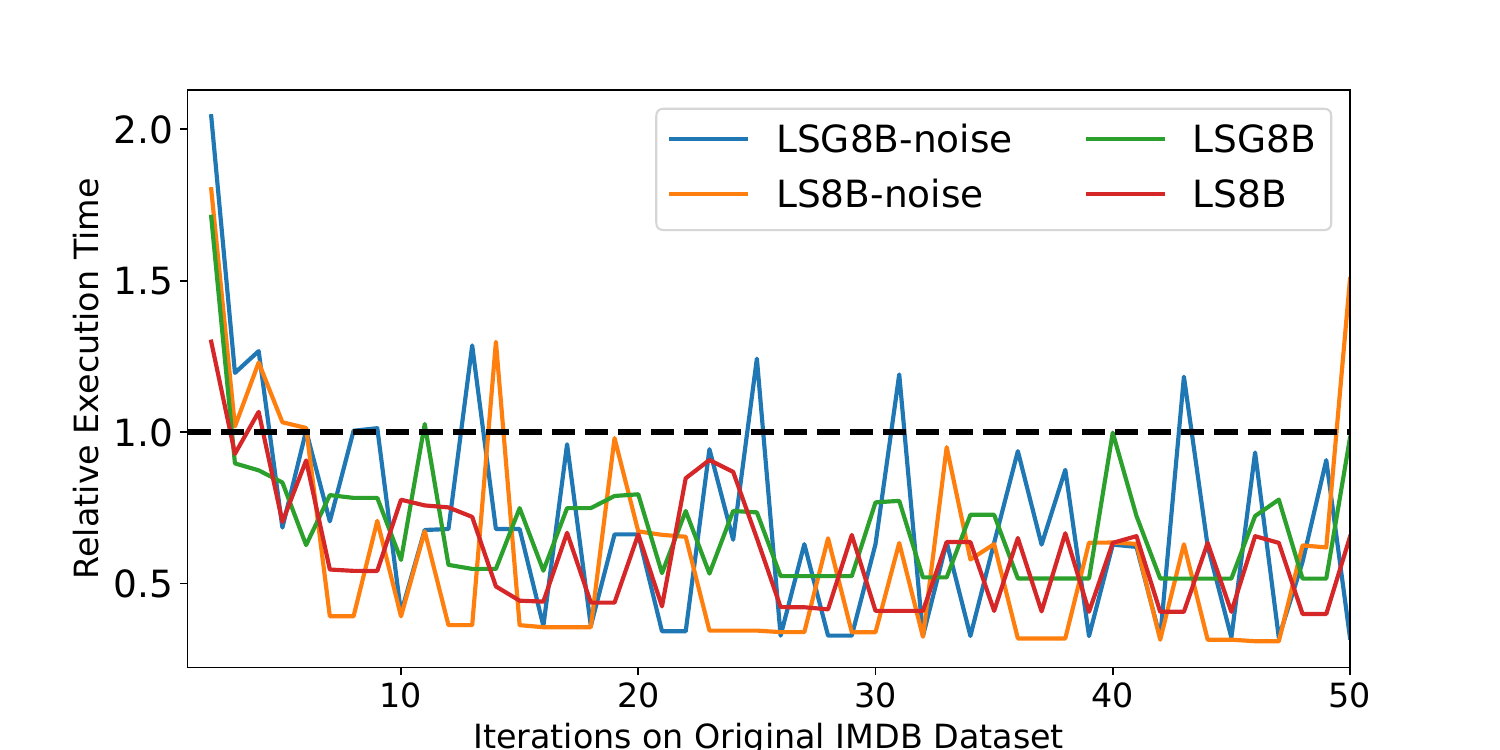}
    \caption{RET for data distribution shifting situations $\downarrow$.}
    \label{fig:data_shift}
\end{figure}

\noindent \textbf{Similar Queries as Reference.}
A key feature of our system is extracting reference queries by similarity search and integrating them into the prompt for in-context learning. This experiment evaluates its impact. Comparing LSG3B with LSG3B-NR, Table~\ref{tab:ablation_studies} reports the performance gain and Figure~\ref{fig:ret_ceb_ablation} presents the \textit{RET}. These results indicate that LSG3B consistently outperforms LSG3B-NR, demonstrating the effectiveness of our design.

\noindent \textbf{How Many References are the Best Setup?} We evaluate {\oursystem} with LS3B on the CEB static workload by varying \(k\) (the number of reference queries in the prompt) from 1 to 3 and assess the impact on performance. Table~\ref{tab:k_influence} presents the performance gains and the lowest \textit{HR} for each configuration. Observing the \textit{HR} metric, we find that as \(k\) increases, more enriched prompts lead the LLM to generate more dynamic hints that deviate from the default reference query plan. However, the performance gain does not improve monotonically with \(k\), indicating that too few or too many reference queries may lead to a suboptimal solution. Too few references leave {\oursystem} lacking key information that could benefit performance, while too many references may confuse the LLM and cause performance degradation.

\begin{table}[htbp]
  \centering
  \caption{Performance gain $\uparrow$ and HR $\downarrow$ for different $k$.}
  \label{tab:k_influence}
  \begin{tabular}{lccc}
    \hline
    $k$ & \multicolumn{2}{c}{Performance Gain} & Optimal HR \\ \cline{2-3}
            & Overall         & Filtered         &    \\ \hline
    1      & 53.93\% & 56.76\%  &  0.26   \\ 
    2     & 57.75\%  & \textbf{62.16\%}  &   0.20 \\ 
    3  & \textbf{57.99\%}           & 58.87\%          &  \textbf{0.14}  \\  \hline
  \end{tabular}
\end{table}

The above ablation studies confirm the crucial role of reference queries in the prompt. They help identify the optimal hyperparameter $k$ and generation mode for {\oursystem}, showing that a moderate prompt length balances inference latency and performance.

\subsection{Extra Investigations}\label{sec:extra-investigation}
\noindent\textbf{Handling Complex Queries.}
To evaluate the capability of {\oursystem} to handle complex workloads, we choose TPC-DS~\cite{tpcds} as a representative benchmark. We replace our local LLMs with GPT-o3-mini~\cite{openai-gpt-o3-mini}, which is expected to deliver better performance due to its larger number of model parameters. {\oursystem} achieves a $13.28\%$ Overall Performance Gain and a $27.81\%$ Filtered Performance Gain. Out of $99$ queries, $28$ have used new query plans from {\oursystem}. Table~\ref{tab:top_5_sql_templates_tpcds} shows the performance gain for the top 5 queries. The query from template $10$ has the highest performance gain at $90.63\%$.

\begin{table}[ht]
\centering
\caption{Top 5 performance gains in TPC-DS queries $\uparrow$.}
\label{tab:top_5_sql_templates_tpcds}
\begin{tabular}{l r c r}
\toprule
Query & Original & Performance Gain & Reduced Time \\
\midrule
10 &  6537.78 ms & 90.63\% & 5925.19 ms \\
17 &  3867.00 ms & 71.79\% & 2776.12 ms \\
25 &  3024.58 ms & 52.59\% & 1590.32 ms \\
16 & 18866.41 ms & 28.80\% & 5433.53 ms \\
79 &   535.55 ms & 27.27\% &  146.04 ms \\
\bottomrule
\end{tabular}
\end{table}

\noindent\textbf{Resilience to Inaccurate Cardinality.}
Cardinality estimation is important for query optimization; inaccurate cardinality could lead to sub-optimal query plans. Therefore, it's necessary to investigate its effect on {\oursystem}. We evaluate the static CEB workload by perturbing each filter and table's cardinality using a random scaling factor~$n$. When the original estimate is $\geq 10$, $n$ is sampled uniformly from $[-1, 1]$; otherwise, from $[0, 1]$.
Note that the database itself remains unchanged; only manual estimation errors are introduced in the prompt. When using LSG8B, {\oursystem} achieves a \(51.50\%\) Overall Performance Gain and a \(57.11\%\) Filtered Performance Gain; when using LS8B, it achieves \(50.26\%\) and \(54.48\%\), respectively. A slight degradation can be obeserverd, however, {\oursystem} is still comparable to the static experiment in Section~\ref{sec:MainResult}.


\noindent\textbf{Robustness under Data Drift.} Here, we study how well {\oursystem} can handle drifts in the underlying data over time. Besides the original IMDB database, we construct a downsized version containing only records dated after the year 2000. We evaluate {\oursystem} on the CEB workload by alternating between these two databases every 5 iterations, and other conditions are just the same as Section~\ref{sec:MainResult}. 
Figure~\ref{fig:data_shift} presents the execution times over 50, showing only the 5 iterations that use the original database in each alternation cycle, to allow comparison with the corresponding static environment. We observe more frequent spikes, indicating that feedback from the downsized distribution interferes with {\oursystem}’s self-evolution process. Nevertheless, for the LS8B model, this variance occasionally leads to the least execution times in some iterations.

\noindent\noindent\textbf{Effect of Different Similarity Calculation Methods.} In {\oursystem}, we employ cosine similarity to retrieve similar queries as in-context learning examples. Besides, we also evaluate inner product and \(L_2\) distance similarity. On the static CEB workload, {\oursystem} with LSG8B using default cosine similarity yields approximately an average of 3\% greater performance gain than the other methods. Interestingly, inner product introduces the highest variance in generated query plans, producing 150 more distinct plans over 50 iterations compared to the other methods.

\noindent\textbf{Can {\oursystem} Operate with Different DBMSes?}
We evaluate {\oursystem} with LSG8B on MySQL using the CEB workload to test whether the knowledge can be transferred between different DBMSes. 
Although these LLMs are supervised fine-tuned on PostgreSQL’s hint format, we apply a simple hint translator (We translate hints from PostgreSQL to MySQL by picking each table in a sequence from the "Leading" clause to generate a join order to change MySQL's behavior) and achieve an Overall Performance Gain of $26.28\%$ and a Filtered Performance Gain of $30.98\%$. These results demonstrate that {\oursystem} can perform effectively in a cross-DBMS environment.

\section{Related Work}


\noindent\textbf{Non-LLM Learned Query Optimizers.} In recent years, Learned Query Optimizers (LQOs) have experienced significant advancements, with a variety of neural architectures—such as Tree-CNNs~\cite{marcus2021bao, yang2022balsa, marcus2019neo}, Tree-LSTMs~\cite{yu2022costlstm, yu2020reinforcementlstm}, and GNNs~\cite{chen2023logergnn, chang2024robustgnn} which are employed to serve as cost predictors or evaluators within these systems. Broadly speaking, LQOs can be categorized based on their generation paradigm into two types: \textit{steering} and \textit{generative} categories. In the \textit{steering} category, the LQO refines query plans generated by traditional optimizers using neural networks. Bao~\cite{marcus2021bao} adjusts the traditional query optimizer behavior by enabling/disabling the different hint sets that this optimizer can choose from. LEON~\cite{chen2023leon} employs a ranking-based approach combined with uncertainty-driven exploration to enhance the query execution performance. LOGER~\cite{chen2023logergnn} leverages a graph transformer along with a beam search strategy to achieve improved execution speedups. Lero~\cite{zhu2023lero} uses a relative-ranking-based method to choose the optimal plan. To sum up, this type has the advantage of being minimally invasive, preserving the core logic of the database engines and reducing the risk of producing poorly performing plans. However, these LQOs are inherently limited by the capabilities of the traditional optimizers they rely on. In the \textit{generative} category, the LQO directly controls the plan generation process. Balsa~\cite{yang2022balsa} and Neo~\cite{marcus2019neo} use the same neural network structure to generate the query plan step by step using reinforcement learning. Neo uses the typical Tree-CNN, and Balsa adds a simulation strategy to boost the performance. Lemo~\cite{mo2023lemo} uses a shared buffer manager component and a plan search algorithm to generate a better plan. These strategies allow for exploring a wider range of execution plans, which can lead to the discovery of more efficient execution. Nevertheless, this greater flexibility also introduces higher variability and potential instability in performance. \\


\noindent\textbf{LLMs and RAG for Databases. }
LLMs demonstrate strong performance across various NLP tasks~\cite{naveed2023comprehensive,questionanswering,codegenerationsurvey}, and are increasingly applied to database tasks such as knob tuning~\cite{db-bert,huang2024llmtune,giannakouris2024demonstrating,lao2023gptuner}, Text-to-SQL~\cite{text2sql-benchmark,text2sql-survey}, code generation~\cite{trummer2025generating}, and schema understanding~\cite{llm-schemata}. DBBert~\cite{db-bert} and GPTuner~\cite{lao2023gptuner} use LLMs to interpret database documents for knob tuning. There is potential to integrate query-level planning and optimization with system-level knob tuning in {\oursystem}, a combination rarely explored in previous work because these two approaches are typically considered orthogonal. CARD~\cite{scholak2021picard} and GPT-DB~\cite{trummer2023demonstrating} leverage LLMs to generate relational data analytics code and SQL pipelines for data analysis. PICARD~\cite{scholak2021picard} further enforces SQL syntax through incremental constraint decoding, while CAESURA~\cite{urban2023caesura} translates natural language into hybrid query plans combining SQL and Python UDFs. 
Similarly, RAG has been widely applied to tasks such as text generation~\cite{guu2020retrievaltextrag,nakano2021webgpttextrag}, visual question answering~\cite{lin2023fvqavisualquestionanswer,chen2022muragvisualquestionanswer}, code synthesis~\cite{lu2022reacccodegen,ding2022cocomiccodegen,codegenerationsurvey}, and recently, it has been used to effectively address many challenges in the database domain. 
For example, in SQL generation, systems like~\cite{shen2024improving} improve accuracy by retrieving structurally similar queries and schema context. Query rewriting also benefits from RAG; approaches like Rafe~\cite{mao2024rafe} retrieve prior rewrite examples to transform natural language queries into efficient, semantically equivalent forms, often surpassing traditional optimizer rules~\cite{sun2024r}. In text-to-SQL, ReFSQL~\cite{zhang2023refsql} reduces hallucinations by retrieving relevant question–SQL pairs to guide generation. 

Recently, three early trials to explore the LLM in the query optimization context, LLMSteer~\cite{akioyamen2024unreasonable}, LLM-QO~\cite{llm4qo-1}, and LLMOpt~\cite{llm4qo-2}, have been proposed. LLMSteer uses only embedding models for hint set selection, which is light, but the improvement is limited. LLM-QO focuses primarily on applying fine-tuning to improve the model's performance. LLM-QO uses a DPO-based reinforcement fine-tuning method, which requires constructing a dataset of ``preferred'' and ``non\_preferred'' pairs beforehand. In contrast, our GRPO-based approach eliminates this overhead and potentially enhances the reasoning abilities of LLMs by training in actual execution environments. On the other hand, LLMOpt~\cite{llm4qo-2} employs LLMs to replace components in the Bao~\cite{marcus2021bao} workflow, fine-tuning them for plan candidate generation and selection. These works represent initial explorations of applying LLMs to replace LQO pipeline. However, {\oursystem} adopts an LLM-native perspective to define optimization problems via model fine-tuning and prompt optimization. Our self‑evolving RAG design also shows continuous learning capabilities that previous methods cannot easily achieve, and SERAG~\cite{liu2025serag} is our initial work.

\section{Conclusion and Future Work}
This paper proposes {\oursystem}, a self-evolving RAG system for query optimization. It leverages LLMs' fine-tuning strategies to avoid cold starts and enhance generalization across workloads. It introduces a two-phase framework that combines offline SFT, {\qgrpo} and online in-context learning. {\oursystem} dynamically generates prompts and continuously evolves through execution feedback. We explore the knowledge transfer capabilities of LLM-based query optimization and demonstrate its potential for practical deployment. Various experiments show that {\oursystem}'s superiority over previous LQOs and the baseline execution engine. To the best of our knowledge, this work is the first to apply RAG to query optimization in databases, thereby extending the frontier of this research area. Currently, {\oursystem} targets OLAP queries, as its overhead remains a challenge for OLTP. Overcoming this limitation is left for future work.


\newpage
\bibliographystyle{plain}
\bibliography{references}
\end{document}